\documentclass[a4paper]{article}

\usepackage{makeidx}         
\usepackage{graphicx}        
\usepackage{multicol}        
\usepackage[bottom]{footmisc}

\usepackage{amsmath, amsfonts, xcolor, comment, verbatim, multirow}
\usepackage{hyperref}
\usepackage{float}

\usepackage{tikz}

\newcommand{\notes}[1]{\textcolor{blue}{[\textsf{#1}]}}
\newcommand{\myparagraph}[1]{\paragraph{#1}}
\newcommand{\covid}{COVID-19}
\newcommand{\sars}{SARS-CoV-2}
\newcommand{\ETI}{\mathrm{ETI}}

\newcommand{\mat}[1]{\mathsf{#1}}
\makeatletter \g@addto@macro\UrlSpecials{\do\!{\newline}}
\renewcommand{\vec}[1]{\mathbf{#1}}

\makeindex             

\begin{document}

\title{Pooled testing and its applications \\ in the \covid\ pandemic}
\author{Matthew Aldridge\** and David Ellis\textsuperscript{\textdagger}}
\date{{\small \**University of Leeds, UK. m.aldridge@leeds.ac.uk \\ \textsuperscript{\textdagger}University of Bristol, UK. david.ellis@bristol.ac.uk}}
\maketitle

\begin{abstract}
When testing for a disease such as \covid, the standard method is individual testing: we take a sample from each individual and test these samples separately. An alternative is pooled testing (or `group testing'), where samples are mixed together in different pools, and those pooled samples are tested. When the prevalence of the disease is low and the accuracy of the test is fairly high, pooled testing strategies can be more efficient than individual testing. In this chapter, we discuss the mathematics of pooled testing and its uses during pandemics, in particular the \covid\ pandemic. We analyse some one- and two-stage pooling strategies under perfect and imperfect tests, and consider the practical issues in the application of such protocols.

This is an extended version of a book chapter to appear in \emph{Pandemics: Insurance and Social Protection}, edited by M. C. Boado-Penas, J. Eisenberg and \c{S}. \c{S}ahin and to be published by Springer.
\end{abstract}


\newpage
\setcounter{tocdepth}{2}
\tableofcontents
\newpage


\section{Introduction}

When testing for a disease such as \covid, the standard method is \emph{individual testing}: we take a sample from each individual and test these samples separately.
Under the convenient mathematical model of perfect testing, a sample from an infected individual always gives a positive result, while a sample from a uninfected individual always gives a negative result. For $N$ individuals, this requires $N$ tests, and we will correctly classify every individual as infected or uninfected. The infected individuals can be advised to self-isolate and their contacts can be traced, while the uninfected individuals are reassured that they are free of the disease.

An alternative to individual testing is \emph{pooled testing}, also called \emph{group testing}. Instead of testing individual samples, we can instead pool samples together and test that pooled sample. Again under the convenient model of perfect testing, a pool consisting entirely of uninfected samples gives a negative result, while a pool containing one or more infected samples gives a positive result. Thus a negative result demonstrates that every individual in the pool is uninfected, while a positive result requires further information to work out which individuals in the pools are infected.

As we shall see in this chapter, when the prevalence of a disease is low enough and the accuracy of a test is high enough, pooled testing can accurately classify individuals as infected or uninfected in fewer than $N$ tests, and sometimes much fewer. This can be more efficient -- and often much more efficient -- than individual testing.


 This chapter is structured as follows. In the remainder of this section, we introduce background material. In Section \ref{sec:perfect-tests}, we analyse some algorithms for pooled testing under an idealised model of perfect tests. In Section \ref{sec:noisy}, we adapt this analysis to more realistic models of testing with errors. In Section \ref{sec:quant}, we briefly discuss a different model called `quantitative pooled testing'. In Section \ref{sec:challenges}, we discuss some practical issues and problems with the application of pooled testing for \covid. In Section \ref{sec:uses}, we survey some uses of pooled testing during the pandemic, so far. In Section \ref{sec:personal-recommendations}, we conclude, and give some of our own views on potential applications of pooled testing for \covid.

\subsection{Testing for \covid}
\label{sec:testing-for-covid}
As well as discussing the general theory of pooled testing, much of this chapter concerns applications of pooled testing in the \covid\ pandemic, so we proceed to give some background on the existing tests for detecting current infection SARS-CoV-2, the virus that causes \covid.

In the real world, testing is not perfect. We distinguish between two types of test errors:
\begin{itemize}
\item \emph{False positive test errors}, where a sample (individual or pool) that does not contain any infection wrongly gives a positive result. The probability that an infection-free sample correctly gives a negative result is called the \emph{specificity}.
\item \emph{False negative test errors}, where a sample (individual or pool) that does contain infection wrongly gives a negative result. The probability that an infected sample correctly gives a positive result is called the \emph{sensitivity}.
\end{itemize}

The most commonly used test for \sars\ infection is the RT-PCR test (reverse transcription polymerase chain reaction test, or just `PCR test' for short). A PCR test for \sars\ infection typically works as follows. First, a swab is taken from the nose or upper throat of the individual to be tested. The swab is then sent to a laboratory, where material from the swab 
analysed to find out whether it contains genetic material from the \sars\ virus. In detail, the sample is first processed to isolate the RNA, then the RNA is then converted to DNA via reverse transcription. Using successive cycles of heating and cooling 
together with enzymes, the relevant DNA is then replicated, with the amount of relevant DNA increasing with each cycle. Fluorescent probes are then inserted, which fluoresce in the presence of a sufficient quantity of the relevant DNA. If RNA from \sars\ was present in the original sample, then after sufficiently many cycles, there will be enough fluorescence to detect. The number of cycles required -- the \emph{cycle threshold}, or CT value -- depends on the quantity of viral RNA present in the original sample, and therefore on the viral load in the individual being tested.
The process typically takes from four to six hours from the receipt of the swab until the output of the result, depending on the laboratory \cite{mahase-length}.

The PCR test is very highly specific, with specificity estimated at ranging from 97.4\% to 99.98\% \cite{skitrall}, meaning that false positive test errors are extremely rare. (The tests used in the UK's ONS Coronavirus Infection Survey, for example, have a specificity of more than 99.92\%; see \cite{ons-meth}.) On the other hand, the PCR test is only moderately sensitive, with sensitivity in the range $70$--$90\%$ being typical \cite{boger,woloshin}. The sensitivity depends on the laboratory protocol being used (such as the maximum number of cycles), and can be affected by shortages of reagent or improper procedures (as can arise in overstretched or under-resourced testing systems).
Another significant source of insensitivity is improperly taken swabs; this can depend on the level of training of the person taking the swabs,
so sensitivity can be lower in community settings than in healthcare settings \cite{bmj-sens}. Sensitivity for a given individual also depends on the viral load and the how long after illness onset the swab was taken. The mathematical lesson from all this is that a negative test does not definitively rule out the individual being infected.

Another test for \sars\ infection is the RT-LAMP (reverse transcription loop-mediated isothermal reaction) test. The studies available so far suggest sensitivity and specificity slightly lower than those of PCR tests \cite{subsoontorn}. Pooled testing can certainly be used with RT-LAMP tests; however, they are not yet widely available, so we will focus our attention here on PCR tests when discussing COVID-related applications of pooled testing.

A third test is the lateral flow test, which in 2021 is being used in the UK for mass-testing in certain areas (such as areas with high prevalence, or where new variants have been detected), and for the regular screening of secondary school pupils (with pupils being asked to self-test at home twice per week, from 15 March 2021). Lateral flow testing kits are cheap, easily portable, require little training to use, and produce a result in around 30 minutes; on the flip side, they have much lower sensitivity than PCR tests. For example, in a large pilot study in the city of Liverpool, the sensitivity of community-based lateral flow testing was estimated at 48.9\% (95\% CI: 33.7\% to 64.2\%) \cite{liverpool-pilot}. We believe that pooled testing is unlikely to be compatible with the lateral flow testing programme in the UK, in view of the low sensitivity of the test, the level of training of those administering the tests, and the premium placed on rapid turnaround time. 


 Much of our analysis in this chapter is applicable to testing for other pandemic diseases, such as pandemic influenza, sometimes with adjustments to the assumed sensitivities and specificities of the tests being used. In particular, the mathematical models used are independent of the disease in question, though the assumptions may be more or less appropriate in the case of other diseases. For example, if very accurate and rapid tests are available for a certain pandemic disease, then pooled testing algorithms with more than two sequential stages may be well worth considering, as they can yield even greater resource-savings than pooled testing algorithms with one or two stages.

\subsection{Stages of a pooled testing algorithm}
\label{sec:stages}

Pooled testing was first proposed in 1943 by Robert Dorfman \cite{dorfman} for the detection of cases of syphilis in those called up for US army service during the Second World War. (The textbook of Du and Hwang \cite[Chapter 1]{du-hwang} gives more information about the early history of pooled testing.) \emph{Dorfman's algorithm} is perhaps the simplest of all pooled testing algorithms, and has also been the most widely-used one in disease control, both prior to and during the \covid\ pandemic. It proceeds as follows. We assume for the moment that tests are perfectly accurate, with $100\%$ sensitivity and $100\%$ specificity.

\begin{figure}[t]
\includegraphics[width=\textwidth]{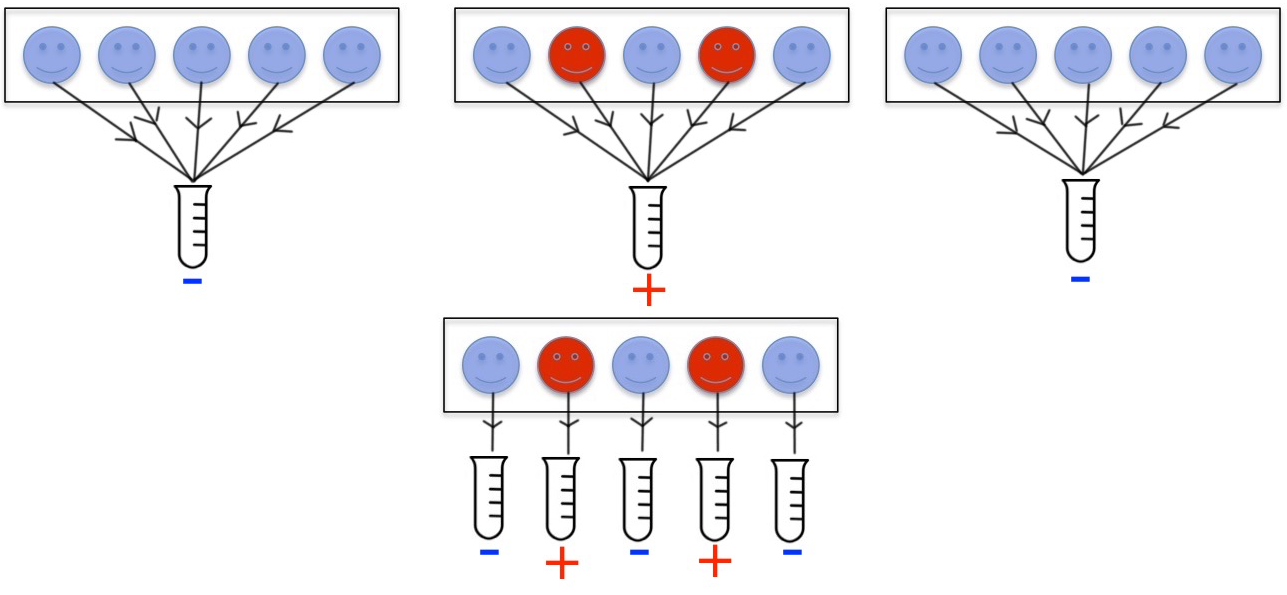}
\caption{Schematic illustration of the use of Dorfman's algorithm (under perfect testing) to identify all the infected individuals in a group of 15, using pools of size 5. In the above case, there are two infected individuals, and only eight tests are required to identify them. If the two infected individuals had been in different pools at the first step, then 13 tests would have been required.}
\label{dorfman-illustration}
\end{figure}

Suppose we have $N$ individuals, and we wish to identify who among those $N$ individuals is infected.
\begin{enumerate}
\item We choose a pool size $s$, and we divide the $N$ individuals into $N/s$ disjoint groups of size $s$ each. (We assume, for simplicity, that $N$ is an exact multiple of $s$.) We take a sample from each of the $N$ individuals, and then, for each of the $N/s$ groups, we pool the $s$ samples from that group into a single pooled sample. We then run a test on each of the $N/s$ pooled samples.
\item 
\begin{enumerate}
\item If a pool tests negative, we know all the individuals in the corresponding group are uninfected.
\item If a pool tests positive, we then follow up by individually testing all the individuals in the corresponding group. These individual tests discover which of the samples in the pool were infected or uninfected.
\end{enumerate}
\end{enumerate}
At the end of this process, under our perfect testing model, we have correctly classified all the individuals as infected or uninfected. This is illustrated in Figure \ref{dorfman-illustration}, in the case $N=15$ individuals and pools of size $s=5$, so there are $N/s = 3$ pools in the first stage.

We shall see later that: 
\begin{itemize}
\item Under perfect testing, if the prevalence is lower than $30\%$, then Dorfman's algorithm uses fewer than $N$ tests on average, so is more efficient than individual testing. (See Subsection \ref{sec:dorf1}.)
\item Under perfect testing, the optimal pool size $s$ is easy to calculate, and is approximately $s = 1/\sqrt{p}$, where $p$ is the {\em prevalence} of the disease (see Subsection \ref{subsec:outline-model} for the formal definition of prevalence), and Dorfman's algorithm requires approximately $2\sqrt{p}N$ tests.
\item Even under imperfect noisy testing, Dorfman's algorithm can be more efficient than individual testing for sufficiently low prevalence. Compared to individual testing, Dorfman's algorithm typically makes many fewer false positive declarations but slightly more false negative declarations. (See Subsection \ref{sec:dorf2}.)
\item For even lower prevalence and higher accuracy, other pooled testing algorithms can not only outperform individual testing but outperform Dorfman's algorithm as well. (See Subsections \ref{sec:grid1} and \ref{sec:rs1}, and \ref{sec:dorf2}.)
\end{itemize}

Note that individual testing is a \emph{one-stage} or \emph{nonadaptive} algorithm, in that all the tests are designed in advance and can be carried out in parallel. Meanwhile, Dorfman's algorithm is a \emph{two-stage} algorithm: the first stage of pooled tests is designed in advance and carried out in parallel, but then the results must be analysed before designing and carrying out the follow-up individual tests in a second stage. 
There are also pooled testing algorithms with more than two stages. There is typically a tradeoff between the number of stages and the efficiency of the algorithm -- more stages allows one to use fewer tests, but more stages take more laboratory time.

It is estimated that approximately 70\% (95\% CI:\ 52\% to 90\%) of the transmission of \covid\ typically takes place either before symptom onset or in the first 48 hours after symptom onset (see \cite{he}, and the small correction in \cite{swiss}). Hence, a fast turnaround time is an important factor to consider when choosing which protocol to use for case detection.
If the tests were to have a very rapid processing time, it might be possible to use algorithms with many stages. However, as stated above, PCR tests for \covid\ typically have a processing-time of four to six hours. It is likely that many laboratories worldwide will be able to perform two sequential stages in a 24-hour period (for logistical reasons, a turnaround time of less than 24 hours from swabbing to result announcement is often hard to achieve anyway), but adding more sequential stages may increase turnaround time too much. Moreover, laboratories under pressure may struggle to keep track of samples over more than two sequential stages. For these reasons, we focus our attention in this chapter on pooled testing algorithms with at most two sequential stages.

For the state of the art in \emph{fully adaptive} algorithms, with no limit on the number of stages, see \cite{aldridge-linear}, as surveyed in \cite[Section 5.5]{aldridge-survey}.

For a more comprehensive (but pre-pandemic) surveys of the mathematics of pooled testing, we refer the reader to the textbook of Du and Hwang (second edition from 2000) \cite{du-hwang}, and a 2019 survey paper by Aldridge, Johnson and Scarlett \cite{aldridge-survey}. 

In this chapter, we mainly only consider pooled testing where each test result is simply either `positive' or `negative'. A different form of pooled testing is where an attempt is made to measure how much viral RNA is present in each pooled sample, and to make use of this information; this is known as \emph{quantitative pooled testing}. We discuss this model briefly in Section \ref{sec:quant}.

\subsection{Who and why to test}

There are two different potential applications of pooled testing for a pandemic diseases such as \covid. The first application, which we have discussed so far, is for {\em case identification}, where it is desired to identify which members of a group are infected, for the purposes of infection-control. There is also a second application, for {\em surveillance}, where the goal is only to estimate the infection prevalence,  without necessarily identifying which individuals are those infected.

In this chapter, we focus mainly on the first application, for case identification, as we believe this is where the most useful applications of pooled testing for COVID-19 are most likely to be found. Briefly, we believe that in the UK, for example, the utility of pooled testing for surveillance on a national scale may be limited in the medium term, because, first, the UK already has a well-developed and extensive national surveillance programmes based on individual testing using random population sampling, such as the ONS Coronavirus Infection Survey, and second, using pooled testing for surveillance only yields large efficiency gains over individual testing when prevalence is lower than it has often been in the UK since the start of the pandemic. Pooled testing for surveillance does, however, still have some potential utility -- for example, if prevalence in the UK becomes sufficiently low and it is desired to reduce the resource requirements of the ONS Infection Survey while still monitoring the prevalence of infection. It is also quite possible that pooled testing could be useful for the surveillance of new variants of the coronavirus, which is important in view of the risks posed by the latter to the effectiveness of vaccination programmes. We return to these issues in Section 6.

 We also draw a distinction between testing symptomatic people, among whom the prevalence is likely to be high, and testing asymptomatic people, where the prevalence is likely to be lower.
 As we shall argue in Section \ref{sec:personal-recommendations}, we believe that pooled testing for case identification is most likely to be useful for the screening of asymptomatic people -- and possibly for the testing of contacts of confirmed cases, provided the prevalence of infection among the group to be tested is thought to be sufficiently low. On the other hand, we believe that pooled testing is unlikely to be useful for the testing of symptomatic people, because the prevalence of COVID-19 infection among those presenting symptoms is usually sufficiently high that the resource savings of pooled testing would be modest compared to individual testing, and are arguably outweighed by the down-sides of pooled testing, such as increased turnaround time compared to individual testing.

\section{Pooled testing algorithms for perfect tests}
\label{sec:perfect-tests}

\subsection{Outline and model}
\label{subsec:outline-model}
In this section, we look at some algorithms for pooled testing for a disease, and assess their performance under the mathematically convenient model of perfect test results.
(Later, in Section \ref{sec:noisy}, we look at the performance of these algorithms in the more realistic model of tests that are highly-but-imperfectly specific and moderately sensitive.)

Unsurprisingly, a key quantity in our model is the {\em prevalence} of the disease, denoted by $p$: this is the fraction of individuals in the population in question who are infected with the disease, at the time when testing is being done. Equivalently, it is the probability that an individual selected at random from the population is infected. 

We assume that $N$ individuals are being tested, and that these individuals are drawn from a large population. Each member of the population is assumed to be infected with probability equal to this prevalence $p$, independently of all other members of the population. (This independence assumption is not quite realistic in many settings, since some of the individuals being tested will often be contacts of one another, so clustering can occur. However, as we shall see, clustering actually makes pooled testing algorithms more efficient than if clustering is not present.)

Assuming that tests are perfect, we usually aim to correctly classify all $N$ individuals as infected or uninfected. We often summarise the performance of algorithms through the \emph{expected tests per individual}.
If an algorithm uses a (possibly random) number of tests $T$ to classify $N$ individuals as either `infected' or `non-infected', then the expected tests per individual is $(\mathbb ET)/ N$, where $\mathbb{E} T$ denotes the expectation (or mean, or average) of the random variable $T$. Clearly, it is desirable for the expected tests per individual to be as small as possible. Note that individual testing clearly has $(\mathbb ET )/ N = N/N = 1$. The expected tests per individual is useful for comparing how much better (or worse) an algorithm is than individual testing.

A standard information-theoretic bound called the \emph{counting bound} (see, for example, \cite{aldridge-linear, aldridge-survey, BJA}) states that, for any successful pooled testing procedure, the expected tests per individual satisfies the lower bound
\begin{equation}
    \label{eq:counting-bound}
\frac{\mathbb ET }{ N} \geq H(p), 
\end{equation}
where $p$ is the prevalence of the disease, and $H(p)$ is the \emph{binary entropy} function, defined by
\[ H(p) = p \log_2 \frac1p + (1-p)\log_2 \frac{1}{1-p}.\]
The bound (\ref{eq:counting-bound}) immediately implies that, when the prevalence is high enough, pooled testing cannot significantly outperform individual testing (under the model of perfect tests). For example, when the prevalence of infection in the population being tested is 25\% ($p = 0.25$), we have $H(0.25) \approx 0.81$, and therefore no pooled testing algorithm can use less than 81\% of the number of tests per individual required by individual testing, at this prevalence level. A deeper mathematical result of Fischer, Klasner and Wegener \cite{indiv} states that (again under the model of perfect tests), individual testing is optimal among all pooled testing algorithms whenever the prevalence is at least 38.2\% ($p \geq 0.382$).

It is useful to consider, for different pooled testing algorithms, how close their `expected tests per individual' is to the counting bound (\ref{eq:counting-bound}), at different prevalence levels, under the model of perfect tests, and this is what we shall do in this section.

\begin{figure}[p]
\includegraphics[width=\textwidth]{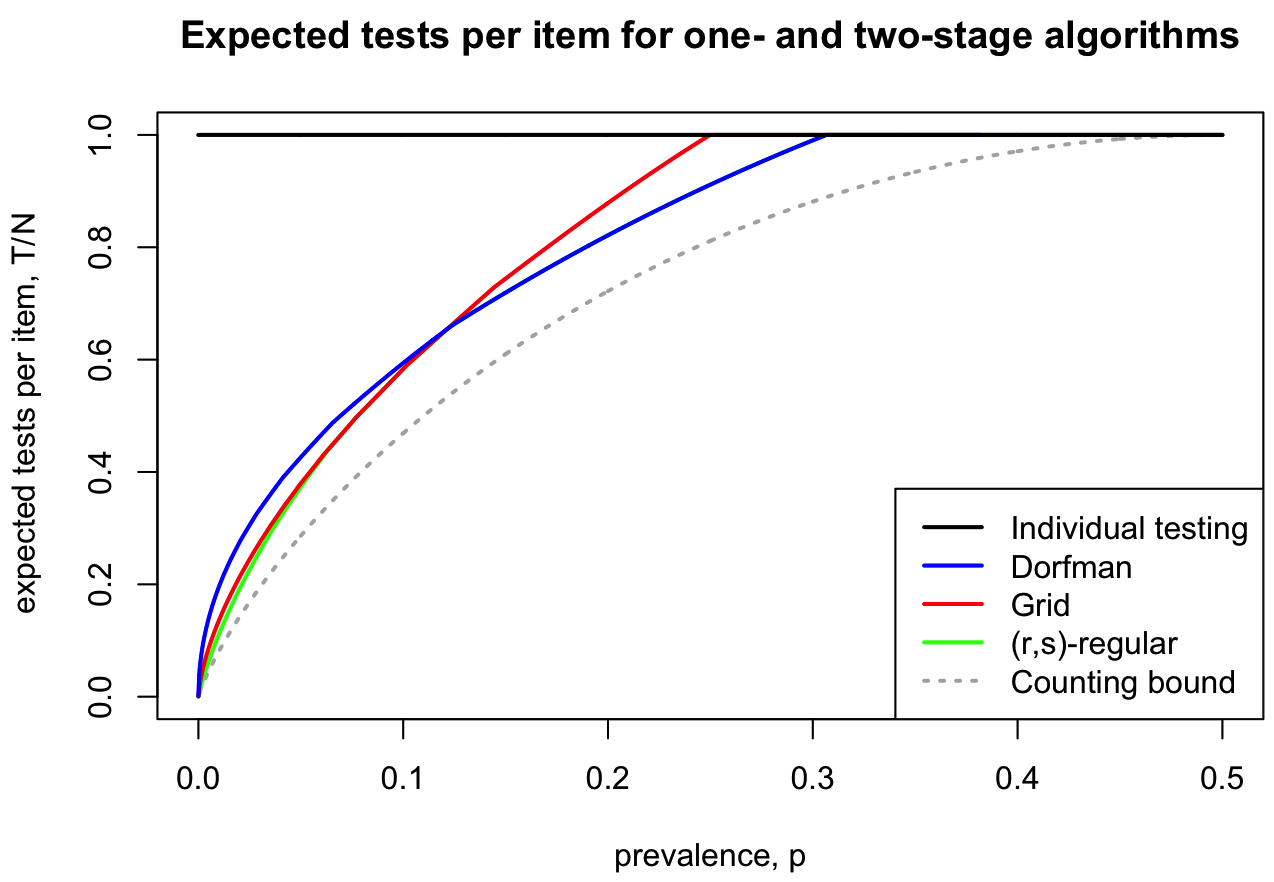}

\vspace{0.6cm}

\includegraphics[width=\textwidth]{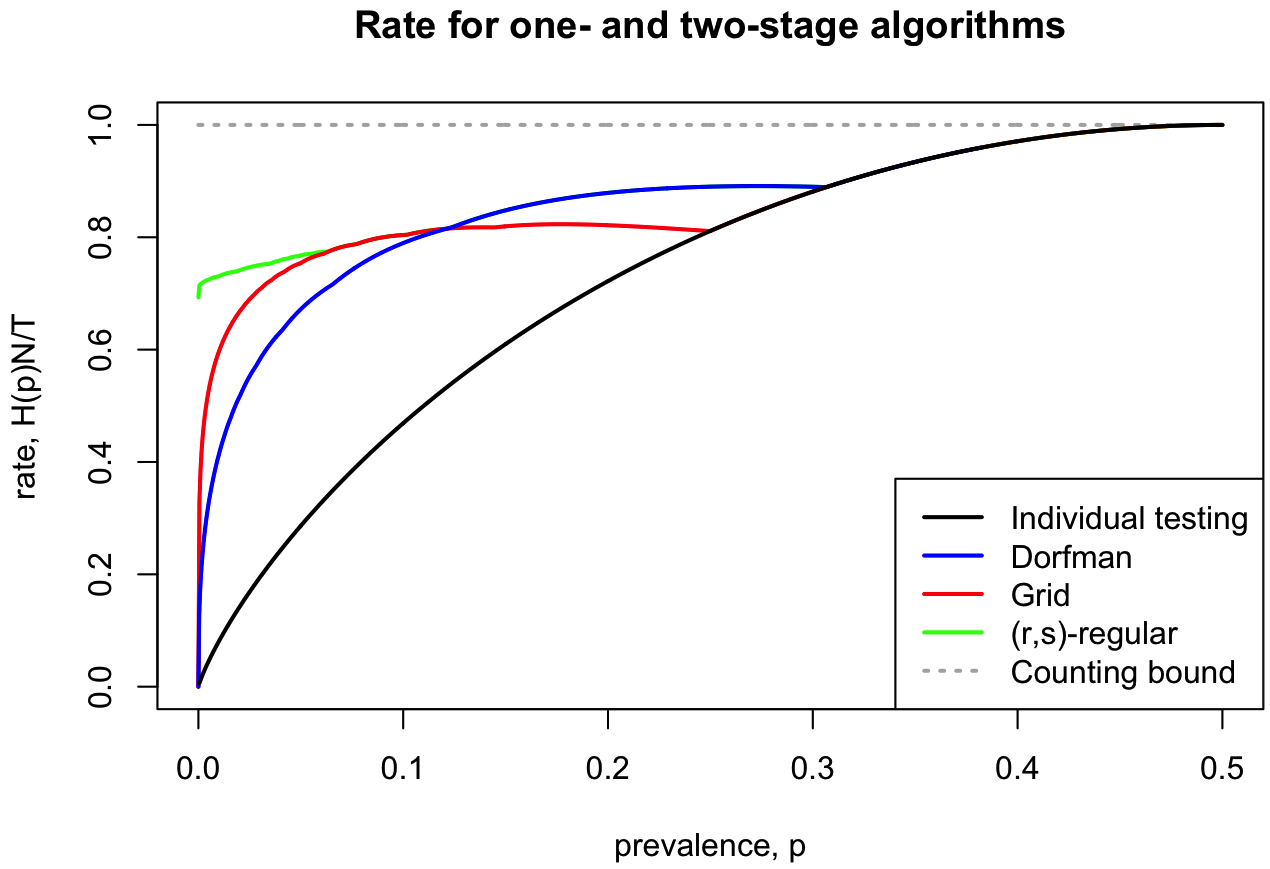}
\caption{Performance of one- and two-stage group testing algorithms under perfect testing, as measured by \textbf{a}~the expected tests per item, and \textbf{b}~the rate. `Grid' and `$(r,s)$-regular' refer to the conservative two-stage variants. At all but the lowest prevalences, the $(r,s)$ regular design has optimal parameter $r =1 $, so is equivalent to Dorfman's algorithm, or $r =2$, and is equivalent to the grid algorithm.}
\label{mainfigs}
\end{figure}

In this section, we study the following four classes of pooled testing algorithms:
\begin{itemize}
    \item individual testing;
    \item Dorfman's algorithm, where each individual's sample appears in exactly one pool in the first stage;
    \item grid-based designs, where each individual's sample appears in exactly two pools in the first stage;
    \item $(r,s)$-regular designs, where each individual's sample appears in exactly $r$ pools in the first stage, with each pool containing $s$ samples.
\end{itemize}

Figure \ref{mainfigs} shows the performance of these algorithms (with the grid and $(r,s)$-regular methods algorithms in what we will later call their `conservative two-stage' variants -- see Subsubsection \ref{subsec:variants-grid}). The top subfigure shows the expected tests per item, and indicates the potential benefit of using pooled testing (compared to individual testing) when the prevalence is below $20\%$.
However, this top subfigure does not sufficiently convey the difference between the pooling algorithms at very low prevalence. The bottom subfigure shows the \emph{rate}, defined by
\[ \text{rate} = \frac{H(p)}{\mathbb ET / N} = \frac{H(p)N}{\mathbb ET}. \]
The rate measures how close an algorithm gets to the counting bound (\ref{eq:counting-bound}) -- higher rates are better. (The rate can also be interpreted information-theoretically, as the number of bits of information learned per test. See, for example, \cite{aldridge-survey, BJA}.) The rate illustrates better the comparison between the different pooling methods, and shows the advantage of the $(r,s)$-regular design at very low prevalences (e.g., below $2\%$).

\subsection{Individual testing}

Clearly, the individual testing of $N$ individuals requires $T = N$ tests, yielding an expected number of tests per individual equal to $T/N = N/N = 1$.

An obvious advantage of individual testing is that one does not need an estimate of the prevalence. It is a one-stage algorithm, so has the fastest possible turnaround time. It is also the simplest testing algorithm to implement. However, we will see that other algorithms yield huge resource savings over individual testing, at low prevalence-levels, and are therefore well worth considering, when prevalence is fairly low and testing capacity is constrained.

\subsection{Dorfman's algorithm} \label{sec:dorf1}


In Section \ref{sec:stages}, we introduced Dorfman's original two-stage algorithm. In this section, we discuss it in more detail.

Suppose we receive samples from $N$ individuals in a large population where the prevalence of infection is $p$, and we run Dorfman's algorithm on these $N$ individuals, using pools of size $s \geq 2$, where $N$ is assumed to be a multiple of $s$. Clearly, one test for each pool is always used at the first pooled testing stage. A pool will also require an extra $s$ individual tests in the second stage if the pooled test was positive. The pooled test will be positive unless all $s$ items are uninfected, so the probability it will test positive is $1-(1-p)^s$. So for each of the $N/s$ pools we definitely have $1$ pooled test, then with probability $1 - (1-p)^s$ another $s$ individual tests are required, giving an expected number of tests as
\[ \mathbb ET = \frac{N}{s} \left(1 + s\big(1 - (1-p)^s\big) \right) = \left(\frac1s + 1 - (1-p)^s \right) N . \]
Hence the expected tests per individual is 
\[ \frac{\mathbb ET}{N} = \frac1s + 1 - (1-p)^s  . \]

It is easy to check that $s = 2$ is never the best choice of pool size $s$, but that $s = 3$ improves on individual testing for $p < 1 - (1/3)^{1/3} = 0.307$. That is, Dorfman's algorithm improves on individual testing for prevalences below roughly $30\%$.

If the prevalence $p$ is known accurately beforehand,
we should choose the pool size $s$ so as to minimise the quantity $1/s + 1-(1-p)^s$, and thereby minimise the expected number of tests required. For fixed $p$, the function $s \mapsto 1/s + 1-(1-p)^s$ is very well-behaved, and it is very easy to numerically find the integer $s$ that minimises it.
But it is useful also to note that when $p$ is small, we may use the approximation
\[ \frac{1}{s} +1-(1-p)^s \approx \frac{1}{s} +1 - (1 -ps) = \frac{1}{s} +ps , \]
which is minimised over the reals at $s = 1/\sqrt{p}$. This gives an expected tests per individual of approximately $2\sqrt{p}$. More formally, choosing $s = \lfloor 1/\sqrt{p} \rfloor$ yields that the expected number of tests per individual is \(1/s +1-(1-p)^{s} = 2\sqrt{p} + O(p)\), where the error term $O(p)$ is small compared to $\sqrt{p}$ when $p$ is small. Hence, when $p$ is small, a good estimate for $p$ is known beforehand, and  the pool-size $s$ is chosen sensibly, our model predicts that Dorfman's algorithm uses, on average, approximately $2\sqrt{p} N$ tests to identify all the infected individuals among a population of size $N$.


Table 1 shows the predicted resource requirements (under the simple model above, i.e., perfect tests), of using Dorfman's algorithm to test $N$ individuals, at different prevalence levels, when the prevalence level is known accurately beforehand, and the pool-size $s$ is chosen optimally. 

\renewcommand{\arraystretch}{1.6}
\setlength{\tabcolsep}{6pt}

\begin{table}[!t]
\caption{Resource requirements of using Dorfman's algorithm to test $N$ individuals, at different prevalence levels, with tests of perfect sensitivity and specificity.}\label{table:sens100}
\begin{center}
\begin{tabular}{ccc}
\hline
\multirow{2}{*}{Prevalence $p$} & \multicolumn{2}{c}{\textbf{Dorfman's algorithm}} \\[-4pt]
 & Optimal pool size $s$ & Expected tests per individual \\
\hline
$5\%$ & $5$ & $0.43$ \\
$2\%$ & $8$ & $0.27$ \\
$1\%$ & $11$ & $0.20$ \\
$0.5\%$ & $15$ & $0.14$ \\
$0.2\%$ & $23$ & $0.088$ \\
$0.1\%$ & $32$ & $0.063$ \\
\hline
\end{tabular}
\end{center}
\end{table}


There are shortcomings in this analysis. We mention three here. Firstly, tests used in the real world do not have perfect sensitivity or perfect specificity; we will refine the model to take this into account, in the following section. 

Secondly, an accurate estimate for the prevalence may not be known beforehand, so it may not be possible to choose the pool size $s$ optimally beforehand. We return to this issue in Section \ref{sec:prevalence-estimates}.

Thirdly, the independence assumption may fail because infections of different individuals being tested by the laboratory are unlikely to be truly independent. But if the overall prevalence remains the same, then Dorfman's algorithm will not perform any worse than the above analysis predicts. In fact, it is advantageous in terms of resource requirements if there is a `clustering' of infected individuals in the same pool. In terms of resource use, the worst case for Dorfman's algorithm is when infected individuals are spread between many pool, and the best-case is when infected individuals are concentrated in few pools.

\subsection{Grid algorithms} \label{sec:grid1}

In Dorfman's algorithm, each individual was tested once in the first stage. In a family of algorithms called  \emph{grid algorithms}, each item is tested twice in the first stage. One variant attempts to classify samples as infected or uninfected just from this single stage, while other variants follow up with a second stage of individual testing.

\subsubsection{Variants of grid algorithms}
\label{subsec:variants-grid}


The grid algorithms always begin as follows. Suppose we have $N$ individuals to test. We split these into $N/s^2$ groups each of $s^2$. (We assume for simplicity that $N$ is a multiple of $s^2$.) 
Let us concentrate on a single group. Each of the $s^2$ individuals is swabbed, and the sample from each swab is divided in two, so it can later be a part of two different sample pools. We now picture the $s^2$ individuals as laid out on a $s \times s$ grid. In the first stage we conduct $2s$ pooled tests: we make one pool from each of the $s$ rows of the grid, and one pool from each of the $s$ columns. 
A PCR test is run on each of these sample pools.

We assume, again, that tests are perfectly accurate. What can we learn from these results of these pooled tests?
\begin{description}
\item[\textbf{Case 0:}] If none of the $s^2$ individuals are infected, then all $2s$ tests will be negative. We can confidently state that all the samples are uninfected.
\item[\textbf{Case 1:}] If exactly one out of the $s^2$ individuals is infected, then exactly one row test and one column test will be positive. We can confidently state that the individual at the intersection of that row and column in the grid is infected and that all the other individuals are uninfected.
\item[\textbf{Case 2:}] If two or more of the $s^2$ individuals are infected, then the test results may be ambiguous. If we are lucky, it could be that all the infected individuals are in the same row or the same column of the grid, in which case they can be identified with complete confidence -- we call this \textbf{Case 2A} -- but more often we cannot be certain exactly which individuals are infected -- we call this \textbf{Case 2B}. 
\end{description}

What should one do after receiving the pooled test results? Here, we briefly look at three possible choices, leading to three different variants of the grid algorithm:
\begin{itemize}
\item \emph{One-stage grid algorithm:} In the one-stage variant, we do not perform any follow-up tests. Any individual that was in at least one negative-testing pool is confidently declared to be uninfected. In Cases 0, 1 or 2A, the remaining individuals (i.e., those who appeared in two positive-testing pools) can be confidently declared to be infected, whereas in Case 2B, the remaining individuals are declared to have unclear results. Running just one single stage has the benefit that a laboratory can quickly process the results, but the downside of sometimes producing inconclusive results. 
\item \emph{Standard two-stage grid algorithm:} In this variant, we can run a second stage of individual testing, to clear up ambiguous results, if there are any. That is, in Cases 0, 1 or 2A, the algorithm is exactly the same as the one-stage grid algorithm above, but in Case 2B, all individuals who were not in at least one negative-testing pool, are given an individual test in the second stage; those who test negative are then declared negative, and those who test positive are declared positive.

\item \emph{Conservative two-stage grid algorithm:} Alternatively,  we can run a second stage where all individuals that appeared in two positive-testing pools are given an individual test -- even in Case 1 or Case 2A, where the infected individuals can be logically determined after the first stage. This has the advantage that every infected individual can be definitely confirmed as infected by the `gold standard' of an individual (non-pooled) PCR test, which might be personally reassuring for the individual or their employer, beneficial when tests are imperfect, or required by regulators. Two-stage algorithms where infection must be confirmed with an individual test are often known in the literature as {\em trivial two-stage algorithms}, but in this chapter we will use the term {\em conservative two-stage algorithm}, as it is more descriptive. Conservative two-stage algorithms can be easier to analyse mathematically, than the standard two-stage ones.
\end{itemize}

In real-life situations, when using the one-stage variant of the grid algorithm, one must make a decision on what to do with individuals with an inconclusive result. One option would be to inform all those individuals they should self-isolate as a precaution; another option would be to individually re-test each of them at a later date (effectively running a two-stage algorithm with a delayed second stage); another option would be to restart the testing from scratch with different grids; a more reckless option would be to inform all the individuals being tested that the test results were inconclusive and that they should continue their lives as if they had tested negative. Which option the laboratory or the regulatory authorities choose may depend on the impact of letting an infection go undetected: if the individual in question is a school pupil or a member of the general community in a mass-testing programme, the impact is likely to be much less than if the individual is a healthcare worker working with highly vulnerable patients or a resident-facing social care worker, for example.

In the next two subsections, we give brief analyses of the one-stage and conservative two-stage variants of the grid algorithm, under our convenient assumption that the tests are perfectly accurate. We summarise these results in Table \ref{table:onegrid}.

\begin{table}[t]
\caption{Resource requirements of using two variants of the grid algorithm at different prevalence levels, with tests of perfect sensitivity and specificity. In the one-stage variant, parameters are chosen such that each grid of $s^2$ individuals is correctly classified with probability at least $0.99$. A dash --- means that the method is worse than individual testing. }\label{table:onegrid}
\begin{center}
\begin{tabular}{ccccc}
\hline
\multirow{2}{*}{Prevalence $p$} & \multicolumn{2}{c}{\textbf{One-stage grid}} & \multicolumn{2}{c}{\textbf{Conservative two-stage grid}} \\[-4pt]
 & Optimal $s$ & Tests per ind. & Optimal $s$ & Exp.\ tests per ind. \\
\hline
$5\%$ & --- & --- & $9$ & $0.38$ \\
$2\%$ & --- & --- & $16$ & $0.21$ \\
$1\%$ & $3$ & $0.67$ & $25$ & $0.14$ \\
$0.5\%$ & $5$ & $0.40$ & $38$ & $0.086$ \\
$0.2\%$ & $8$ & $0.25$ & $68$ & $0.047$ \\
$0.1\%$ & $11$ & $0.18$ & $106$ & $0.030$ \\
\hline
\end{tabular}
\end{center}
\end{table}

\subsubsection{An analysis of the one-stage grid algorithm}

We now analyse the one-stage variant of the grid algorithm, the assumption of perfectly accurate tests.

Recall that if there is at most one infected individual in the group of $s^2$ individuals (and provided each PCR test was perfectly accurate) we will know exactly which individual is infected. We want to make a choice of $s$, depending on an estimate of the prevalence $p$. Note that we perform $2s$ tests on $s^2$ individuals, so the number of tests per individual is $2s/s^2 = 2/s$. Thus, to keep the number of tests per individual low, we would like to choose $s$ to be large. But the larger we choose $s$, the more likely it is that our group of $s^2$ individuals will contain two or more infected individuals, leading to (potentially) inconclusive results.


Suppose that the prevalence of infection 
is $p$, and that we wish to choose the parameter $s$ in the grid algorithm in such a way that the algorithm correctly identifies all infected individuals in each group with probability at least $0.99$. The one-stage grid algorithm successfully identifies all the infected individuals provided all the infected individuals share the same row, or all share the same column (or there are no infected individuals at all). In particular, it certainly succeeds if there are exactly zero or one infected individuals in the grid of $s^2$ individuals. By a union bound, the probability that there are two or more infected individuals is at most
\[ \binom{s^2}{2} p^2 = \tfrac{1}{2}s^2(s^2-1)p^2 \leq \tfrac{1}{2}s^4 p^2. \]
This is no more than $0.01$ provided
\[ s \leq \left(\frac{2 \times 0.01}{p^2}\right)^{1/4} = \frac{(0.02)^{1/4}}{\sqrt{p}} = \frac{0.376}{\sqrt{p}} . \]
Hence, the grid algorithm correctly identifies all the infected individuals in each grid with probability at least $0.99$ provided $s \leq 0.376/\sqrt{p}$. To minimise the expected tests per individual subject to this 0.99 condition, we should clearly choose $s$ to be $\lfloor 0.376/\sqrt{p}\rfloor$, the largest integer less than or equal to $0.376/\sqrt{p}$.

Table \ref{table:onegrid} above showed the results for various values of the prevalence $p$ with the $0.99$ accuracy guarantee. For example, when the infection prevalence $p = 0.2\%$ (the estimate of the ONS Infection Survey for prevalence in the community in England, the week ending 19 September 2020 \cite{ons-19sept}), the above translates to $s \leq 8.41$. Choosing $s=8$, the grid algorithm requires $2s = 16$ tests for testing $s^2 = 64$ people. We obtain $\mathbb{E}T/N = 2/s = 1/4$, which is four times the efficiency of individual testing. However, this is a lot worse than the conservative two-stage variant, which has $0.047$ tests per individual, or $21$ times the efficiency of individual testing. This would suggest (under the assumption of perfect tests) that we should only use the one-stage variant of the grid algorithm in this form (with the 0.99 condition), when a second stage of testing would be prohibitively inconvenient and the prevalence is low.

\subsubsection{An analysis of the conservative two-stage grid algorithm}

Recall that in the conservative two-stage variant of the grid algorithm, every individual in two positive pooled tests receives an individual test. Our analysis here is similar to that in \cite{aldridge-trivial,google}.

For each of the $N/s^2$ grids of $s^2$ individuals, there are $2s$ pooled tests in the first stage. An individual can then receive an individual test for one of two reasons. One is that the individual is infected, which occurs with probability $p$. The second is that the individual is uninfected (which occurs with probability $q=1-p$), and that further, its row contains an infected individual (this happens with probability $1 - q^{s-1}$), and its column also contains an infected individual (this also happens with probability $1 - q^{s-1}$).

Hence, using the linearity of expectation, the expected number of tests is
\[ \mathbb ET = \frac{N}{s^2} \big(2s + s^2(p + q(1 - q^{s-1})^2 )\big) = N \left( \frac{2}{s} + p + q(1-q^{s-1})^2 \right) , \]
and the expected tests per individual is \[\frac{\mathbb{E}T}{N} = \frac{2}{s} + p + q(1-q^{s-1})^2 . \]

As before, given the prevalence $p$, this is a well-behaved function of $s$, so it is easy to numerically choose the optimal $s$. Results were shown in Table \ref{table:onegrid} above, and far outperform the one-stage variant when a second stage is available. Comparing with Table \ref{table:sens100}, we see that the conservative two-stage grid algorithm also outperforms Dorfman's algorithm for the values of $p$ under consideration, though not by more than a factor of approximately two, for these values of $p$. We note, however, that the optimal choice of $s$ (the pool-size in the conservative two-stage grid algorithm) for $p=0.1\%$, is 106, and large pool-sizes do have practical down-sides (usually requiring automation, reducing sensitivity, and posing regulatory problems - see Section \ref{sec:challenges}).

From a more mathematical perspective, a similar approximation to that for Dorfman's algorithm shows that, for small $p$, the optimal $s$ is of the order $1/p^{2/3}$; this yields an expected number of tests per individual of the order $p^{2/3}$. For $p$ sufficiently small, this is an improvement on Dorfman's algorithm (where the expected number of tests per individual is of the order $p^{1/2}$). 

\subsection{Pooling algorithms based on $(r,s)$-regular designs} \label{sec:rs1}

The $(r,s)$-regular designs are a family of algorithms that generalise the algorithms we have seen so far. These algorithms have a first stage where each individual's sample appears in exactly $r$ different pools, and each pool contains samples from exactly $s$ individuals. The second stage (if there is a second stage) consists of individual testing -- as with the grid algorithm discussed in the previous section, in the `standard' variant, an individual is given an individual test at the second stage only when their infection status cannot be determined from the first stage, whereas in the `trivial' variant, an individual is given an individual test at the second stage whenever all the pools they appear in test positive at the first stage.

Individual testing is a special case of an $(r,s)$-regular design, with $r = 1$, $s = 1$ and a single stage. Dorfman's algorithm is a special case with $r = 1$, $s > 1$, with two stages. The grid algorithms we discussed above are special cases with $r = 2$.

There are a number of ways to construct a pooling design that is $(r,s)$ regular.
\begin{itemize}
\item \emph{Randomly:} Given a number of individuals $N$, a testing procedure that tests each individual in $r$ pools with each pool consisting of $s$ samples can be chosen uniformly at random from all such procedures. This is easy to do computationally, and convenient for proving mathematical statements. However, the random choice means that rare bad designs are possible, and the lack of structure can make it awkward to carry out in a laboratory setting.
\item \emph{Hypercube:} This method generalises the grid algorithm to higher dimensions. It is required that $s = a^{r-1}$ for some positive integer $a$. Assume that $N$ is a multiple of $a^r$. We split the $N$ individuals into groups of size $a^r$, and we focus our attention on just one group. Imagine that those $a^r$ individuals are placed on an $r$-dimensional $a \times a \times \cdots \times a$ hypercube. Each pool corresponds to an $(r-1)$-dimensional slice of this hypercube, containing $a^{r-1} = s$ individuals. Note that each individual is sampled in $r$ pools, one for each of the $r$ slice directions. Taking $r = 2$, we obtain the grid algorithm. The structure of the hypercube 
can be convenient for implementation, although automation is usually required for pooling the samples, and for $r \geq 3$, the conditions give a somewhat restricted set of possibilities for $s$.
\item \emph{Code-based:} A classical construction of Kautz and Singleton shows how to construct an $(r,s)$-regular design from an error-correcting linear code with appropriate parameters. We point readers to \cite[Section 5.7]{aldridge-survey} or Kautz and Singleton's original work \cite{kautz-singleton} for further details. The extra structure often gives good performance when $N$ is small, although for some values of $r$ and $s$ it is not possible to find a code with appropriate parameters.
\end{itemize}
We note that, by counting the number of times a sample appears in a pool in two different ways, the number of pooled tests $T_1$ used by the first stage of an $(r,s)$-regular design satisfies $Nr = T_1s$, and therefore the number of tests per individual used by the first-stage of an $(r,s)$-regular design is $T_1/N = r/s$.

As stated above, an $(r,s)$-regular pooled testing algorithm can be used in the form of a one-stage, a `standard' two-stage, or a conservative two-stage algorithm, just as with the grid algorithm.

We briefly present a summary of an analysis of the conservative two-stage variant, following \cite{aldridge-trivial,google}. With this variant, any individual whose $r$ stage-one pooled tests are all positive receives and individual test in the second stage. For large $N$, the expected tests per individual in the random $(r,s)$-design (described above) satisfies the following with high probability:
\begin{equation}
\frac{\mathbb ET}{N} \sim \frac{r}{s} + p + q (1 - q^{s-1})^r , \label{rstn}
\end{equation}
where $p$ is the prevalence and $q = 1-p$. Here, $r/s$ is the number of tests per individual in the first stage. An individual requires retesting in the second stage either if it is infected, with probability $p$, or if it is uninfected, with probability $q$, but all $r$ of its tests are positive, each of which happens with probability $1 - q^{s-1}$. Thus if the results of the tests containing a given uninfected individual were independent, then \eqref{rstn} would hold exactly. It turns out that a randomly sampled $(r,s)$-design satisfies this independence condition for most individuals (with high probability); in fact, with positive probability, it satisfies the condition for all individuals.

Further, a lower bound is given in \cite{aldridge-trivial}, which shows that, among all conservative two-stage algorithms, the random $(r,s)$-regular design is extremely close to optimal for all $p < 0.3$.

\begin{table}[!t]
\caption{Resource requirements of using an $(r,s)$-regular design in a conservative two-stage algorithm to test $N$ individuals, at different prevalence levels, with tests of perfect sensitivity and specificity.}\label{table:rs}
\begin{center}
\begin{tabular}{ccc}
\hline
\multirow{2}{*}{Prevalence $p$} & \multicolumn{2}{c}{\textbf{Conservative two-stage $(r,s)$-regular algorithm}} \\[-4pt]
 & Optimal parameters $(r,s)$ & Expected tests per individual \\
\hline
$5\%$ & $(3,13)$ & $0.37$ \\
$2\%$ & $(4,31)$ & $0.19$ \\
$1\%$ & $(5,63)$ & $0.11$ \\
$0.5\%$ & $(7,147)$ & $0.063$ \\
$0.2\%$ & $(8,351)$ & $0.029$ \\
$0.1\%$ & $(9,700)$ & $0.016$ \\
\hline
\end{tabular}
\end{center}
\end{table}

Table \ref{table:rs} shows the performance of the $(r,s)$-regular design according to \eqref{rstn} with an optimal choice of $r$ and $s$. We note that the $(r,s)$-regular algorithm outperforms individual testing, Dorfman's algorithm, and the grid algorithms for all values of $p$ in the table.

Table \ref{table:rs} shows results with the mathematically optimal choice of $(r, s)$, but as the prevalence gets small, these parameter choices can get quite large. This could be unwieldy or even infeasible for a laboratory to carry out, and large values of $s$ (the pool-size) provoke worries about sensitivity (with imperfectly sensitive tests). However, typically $r$ can be reduced somewhat and $s$ reduced quite a lot with only a marginal reduction in performance. For example, at $p = 0.5\%$, the optimal choice is $r = 7$, $s = 147$, giving an expected tests per individual of $0.063$. But reducing the parameters to the much more manageable $r = 3$, $s = 62$ still gives an expected tests per individual of $0.072$, which is only slightly worse. The practically best choice of parameters will depend on a laboratory's capability for carrying out complicated procedures, and worries about the impact of dilution on test sensitivity (see Section \ref{sec:dilution}).

\section{Pooled testing algorithms for imperfect tests} \label{sec:noisy}

\subsection{The model}

In this section, we refine the model of the previous section to take into account the fact that the tests we are dealing with do not always give the correct answer.
Recall that the PCR test has very high sensitivity, typically higher than $99\%$, meaning false positive test results are extremely rare, and has moderate sensitivity, typically between $70\%$ and $90\%$, meaning that false negative results are not uncommon.

Here, we use a very simple model for such tests. We assume that each test on a pool containing at least one infected sample has a fixed probability $u$ of correctly returning a positive result, and that each test on a pool of entirely uninfected samples has a fixed probability $v$ of correctly returning a negative result, independently of the outcomes of all other tests (including of tests on overlapping pools), and independently of the size of the pool (that is, with no `dilution' effect).

Whether or not this is a realistic model will depend upon the main sources of false negatives and of false positives, and therefore on the precise protocol being used and the practical situation. If the main source of insensitivity or nonspecificity is a shortage of reagents, faulty equipment, or faulty lab-procedures, then it is probably quite realistic. If individuals are frequently swabbed incorrectly (as can happen when individuals are asked to self-swab), then incorrectly taken swabs will be an important source of insensitivity, and in this case, unless individuals are re-swabbed at each successive stage of a group-testing algorithm and there are no overlapping pools at any single step, the independence assumption will not be valid. 
Moreover, the assumption that dilution (where a small number of positive samples are diluted by a large number of negative samples), does not affect sensitivity, is likely to be fairly realistic with pool-sizes of 10 or less and with typical viral loads, but will be less realistic with pool-sizes of 100 or more. See Section \ref{sec:challenges} for a further discussion of this issue.

In the rest of this section, we look at some results regarding two-stage and one-stage algorithms, under this model for noisy tests.

\subsection{Analysis of individual testing and Dorfman's algorithm} \label{sec:dorf2}



For a given algorithm, there are (at least) three things we want to know:
First, how many tests do we expect to use? Second, how many false negative declarations do we expect to make? Third, how many false positive declarations do we expect to make? A useful quantity for comparing algorithms is the {\em expected number of tests per isolated individual} (ETI): that is, the expected number of tests used, divided by the expected number of infected individuals correctly discovered and instructed to isolate. Since the isolation of infected individuals is the main public-health goal of a screening programme, ETI is a good measure of how much benefit we are getting per test used (though it does not take into account turnaround time). 

Let us start with individual testing. To test $N$ individuals, this requires exactly $N$ tests. There are $pN$ infected individuals on average, and we find each one if its test correctly gives a positive result, which happens with probability $u$. So on average we correctly find $upN$ infected individuals but falsely miss $(1-u)pN$ of them; so the ETI is $up$. Similarly, also on average, of the $(1-p)N$ uninfected individuals, we correctly identify $v(1-p)N$ of them, but falsely declare $(1-v)(1-p)N$ of them to be infected.

Now consider using Dorfman's algorithm to test $N$ individuals with pools of size $s$, with $N$ a multiple of $s$, and suppose we use the protocol of declaring an individual to be infected only if both their pooled test and their individual test are positive. (If a pool tests positive in the first stage but all the corresponding individual tests in the second stage are negative, the pooled test is assumed to be a false positive.)

First, an individual will make it through to the second stage if either the pool is infected, and correctly gives a positive result; or if the pool is uninfected, but incorrectly gives a positive result. Thus the expected tests per individual is
\[ \frac{\mathbb E T}{N} = \frac{1}{s} + u(1-q^s) + (1-v)q^s , \]
where $q = 1-p$. Here, $1/s$ represents the requirements of the first stage (i.e., the pooled tests), $u(1-q^s)$ represents the requirements of the second stage in the case of a true positive pool result, and $(1-v)q^s$ represents the requirements of the second stage in the case of a false positive pool result. For small $p$, it turns out that an essentially optimal choice for minimising this quantity is $s = \lfloor 1/\sqrt{up} \rfloor$; this can be shown in a similar way to in the previous section. This yields an expected number of tests per individual which is approximately 
$2\sqrt{up} + (1-v)$, compared to $1$ for individual testing. For $p = 0.02$, $u = 0.8$, $v = 0.995$, we get an improvement from $1$ test per individual to $0.25$ tests per individual.

Second, the expected number of infected individuals found is $u^2pN$, as there are $pN$ infected individuals on average, and they are found if both their pooled test and their individual test are correctly positive. The other (on average) $(1-u^2)pN$ infected individuals get false negative declarations. Compared to individual testing, where the total expected number of false negatives is simply $(1-u)pN$, we have
\[ \frac{(1-u^2)pN}{(1-u)pN} = 1+u \leq 2 , \]
and therefore the expected number of false negatives under Dorfman's algorithm can never be more than twice that when individual testing is used.

Third, a uninfected individual is falsely declared infected if both their pooled test is positive -- either due to a false positive test or the presence of an infected individual and a true positive test -- and their individual test is a false positive. This event has probability
\[ \big((1-q^{s-1})u + q^{s-1} (1-v)\big)(1-v) . \]
Hence, the expected number of false positives is
\[ \big((1-q^{s-1})u + q^{s-1} (1-v)\big)(1-v)qN. \]
Again, for small $p$ with $s = \lfloor 1/\sqrt{up} \rfloor$, this is approximately \((\sqrt{pu} + 1 - v)(1-v)qN\). Compared to the expected number of false positives under individual testing, which is $(1-v)qN$, Dorfman gives an improvement by a factor of
$\sqrt{up} + 1 - v$. For $p = 0.02$, $u = 0.8$, $v = 0.995$, Dorfman gives an expected number of false positives which is approximately $0.12$ times its value under individual testing.
Thus Dorfman produces far fewer false positives than individual testing, a feature which is common to many other pooled testing algorithms.

\begin{table}[tp]
\label{table:sens80}
\caption{Expected resource requirements and impact of using individual testing or Dorfman's algorithm to test $N$ individuals, at different prevalence levels, with tests of sensitivity 80\% and specificity 99.5\%.} \label{tab:noisy}
\begin{center}
\begin{tabular}{cccc}
\hline
\multirow{2}{*}{Prevalence $p$} & \multicolumn{3}{c}{\textbf{Individual testing}}  \\[-4pt]
 & $\mathbb E\,\#$ tests & $\mathbb E\,\#$ false neg & $\mathbb E\,\#$ false pos \\
\hline
$5\%$ & $N$ & $0.01N$ & $0.005N$ \\
$2\%$ & $N$ & $0.004N$ & $0.005N$ \\
$1\%$ & $N$ & $0.002N$ & $0.005N$ \\
$0.5\%$ & $N$ & $0.001N$ & $0.005N$ \\
$0.2\%$ & $N$ & $0.0004N$ & $0.005N$ \\
$0.1\%$ & $N$ & $0.0002N$ & $0.005N$ \\
\hline
\end{tabular}

\vspace{0.6cm}

\begin{tabular}{ccccc}
\hline
\multirow{2}{*}{Prevalence $p$} & \multicolumn{4}{c}{\textbf{Dorfman's algorithm}} \\[-4pt]
 & optimal $s$ & $\mathbb E\,\#$ tests & $\mathbb E\,\#$ false neg & $\mathbb E\,\#$ false pos \\
\hline
$5\%$ & $6$ & $0.38N$ & $0.02N$ & $0.0009N$ \\
$2\%$ & $9$ & $0.25N$ & $0.007N$ & $0.0006N$ \\
$1\%$ & $12$ & $0.18N$ & $0.004N$ & $0.0004N$ \\
$0.5\%$ & $17$ & $0.13N$ & $0.002N$ & $0.0003N$ \\
$0.2\%$ & $26$ & $0.08N$ & $0.0007N$ & $0.0002N$ \\
$0.1\%$ & $36$ & $0.06N$ & $0.0004N$ & $0.0002N$ \\
\hline
\end{tabular}

\end{center}
\end{table}
\setlength{\tabcolsep}{6pt}

Under the assumption that the sensitivity $u$ of each test is 0.8 and the specificity $v$ of each test is 0.995, 
Table \ref{tab:noisy} summarises, for different prevalences, the expected number of tests, false negatives, and false positives for individual testing and for Dorfman's algorithm. Note that, compared to individual testing, Dorfman's algorithm dramatically decreases the number of tests required and the number of false positives, but roughly doubles the number of false negatives.

The ETI of Dorfman's algorithm (used in the way above), is
\[ \frac{\frac{1}{s} + u(1-q^s) + (1-v)q^s)}{u^2 p} ; \]
if $p$ is small and we choose $s = \lfloor 1/\sqrt{up} \rfloor$, as described above, then this is approximately
\[ \frac{2\sqrt{up} +(1-v)}{u^2 p} . \]


More general $(r,s)$-regular designs are studied in \cite{aldridge-conference} under the greater simplification that $v = 1$; that is, of perfect specificity. (Recall that typically $v$ can be in excess of $90\%$.) Here there are no false positives, and it is shown that the ETI is
\[ \ETI = \frac{\frac rs + p + (1 - q^{s-1})^ru^r}{u^{r+1}p} . \]
Here the numerator is as in \eqref{rstn} with the extra factor if $u^r$, since all $r$ tests must correctly be positive for an individual to qualify for the second stage. The denominator arises because an individual is infected with probability $p$, and is identified if all $r$ pooled tests and the one individual test are correctly positive.

\subsection{One-stage testing} \label{onestage}

We now briefly consider general one-stage (nonadaptive) pooling algorithms, under our simple model of imperfect tests. Here, we see the results of the tests, and we must try to come up with a `best guess', from those results, as to which individuals were infected. The precise meaning of `best guess' depends (for example) on the down-sides of missing infected cases (false negatives), and on the down-sides of false positives. This kind of problem is known as an {\em inference problem}.

We suppose the pooling design is chosen according to a \emph{pooling matrix} $\mat A = (a_{ti}) \in \{0,1\}^{T \times N}$, a matrix of zeros and ones, where $a_{ti} = 1$ if the sample from individual $i$ is included in the $t$th pooled test, and $a_{ti} = 0$ otherwise. The $T$ rows of the matrix $\mat A$ represent the $T$ pooled tests, and the $N$ columns represent the $N$ individuals being tested.

Some further notation is useful. Let $\vec x = (x_i) \in \{0,1\}^N$ be the vector of zeros and ones where $x_i = 1$ if individual $i$ is infected, and $x_i = 0$ if individual $i$ is uninfected; the vector $\vec x$ represents which individuals are truly infected and which are not, and it is what we really want to guess; we refer to it as the `infection vector'.

We write $\vec y = (y_t) \in \{0,1\}^T$ for the actual outcomes of the tests, i.e., $y_t = 1$ if the $t$th pool tests positive, and $y_t = 0$ otherwise. Finally, we write $\tilde{\vec y} = \tilde{\vec y}(\mat A, \mathbf x)$ for what the $T$ outcomes of the pooled tests on the infection vector $\vec x$ would have been under perfect testing; i.e., $\tilde{y}_t = 1$ if the $t$th pool would have tested positive under perfect testing, and $\tilde{y}_t=0$ otherwise. Explicitly, $\tilde{y}_t=1$ if $(\mat A \vec{x})_t  \geq 1$, and $\tilde{y}_t = 0$ if $(\mat A \vec{x})_t = 0$.

Given $\vec y$ and $\mat A$, we must come up with an estimate $\vec{\hat{x}}$ for $\vec x$. One way to do this is to choose $\vec{\hat{x}}$ so that $\tilde{\vec y}(A,\vec{\hat{x}})$ minimises the `penalty function'
\[ f(\mathbf{\hat{x}}) = a\times\#\{i : \hat{x}_i = 1\}
  + b\times \#\{t : y_t = 1, \tilde y_t = 0\}
  +c\times\#\{t : y_t = 0, \tilde y_t = 1\},
\]
for some choice of constants $a, b, c \geq 0$. Here, $b > 0$ penalises assumed false positive test results, $c > 0$ penalises assumed false negative test results, and $a > 0$ promotes a sparse solution (which is mathematically desirable, if the prevalence is assumed to be low).

In practice, performing a global minimization may be too computationally demanding, depending upon the numbers of individuals involved and the computational resources available, and in this context, a good method (adopted e.g.\ by the P-BEST algorithm, see Section \ref{sec:israel}) is to assume that individuals appearing in `many' negative pools are uninfected (for an appropriate definition of `many'), and then to perform a minimization over all subsets of the remaining individuals.

If we simply wish to find the most likely set of infected individuals (that is, we do not want to err on the side of caution when we report to the individuals whether they are infected or not), then it makes sense to report a maximum {\em a posteriori} (MAP) estimate for $\mathbf x$. This can be found by minimizing a penalty function of the above form, for an appropriate choice of $a, b, c$ (a choice that depends on the assumed prevalence, and the assumed sensitivity and specificity of the tests). The P-BEST algorithm (see Section \ref{sec:israel}) attempts to minimize a simpler penalty function of the above form, with $a = 0$ and $b = c = 1$.

There are other reasonable alternatives for methods of choosing $\tilde{\vec{x}}$, depending upon one's ultimate goal. For example, one might wish to minimise the expected number of false positive declarations and the number of false negative declarations -- perhaps weighted differently, to account for the estimated cost of declaring an infected person uninfected being potentially much greater than the estimated cost of declaring an uninfected person to be infected.

\section{Quantitative pooled testing} \label{sec:quant}

So far in this chapter, we have been considering pooled testing where each test-result (on a pooled sample) is classified only as `positive' or `negative', and where we seek to classify each individual only as `infected' or `uninfected'. In other words, no attempt is made to measure the concentration of viral RNA in any of the pooled samples, or to estimate the viral load in any of the individuals being tested. Alternatively, a different form of testing could be used, where we do measure how much viral RNA is present in the pooled samples, and make use of this extra information. This is known as \emph{quantitative pooled testing}, or \emph{quantitative group testing}.
Quantitative pooled testing algorithms take into account the fact that different pools can contain different concentrations of viral material, depending both on how many individuals in the pool were infected, and on the viral loads present in the samples from each of those infected individuals.

Quantitative pooled testing can have significant advantages over standard (non-quantitaive) pooled testing. Firstly, the measurement of  `total viral load' in a pooled sample is more informative than a simple positive or negative result, so the counting bound \eqref{eq:counting-bound} no longer applies, and there is the potential for greater reductions in the number of tests than a standard pooled testing protocol could achieve. Secondly, quantitative pooled testing allows the estimation of the viral load in an infected individual, which correlates with illness severity and also with infectiousness, and therefore has clinical implications, and applications for infection control. A disadvantage of quantitative pooled testing algorithms is that the measurement of total viral load is more technically challenging and costly than measuring a simple positive or negative result, and may be more time-consuming (although in a well-run laboratory with the appropriate expertise and software, the extra time required would probably be minimal). Quantitative pooled testing could also require significant changes to laboratory workflows, and the retraining of laboratory staff. Another problem is that regulators may be less likely to approve quantitative pooled testing for COVID-19, as the inference of who is infected and who is not is less transparent than with ordinary pooled testing, particularly for those quantitative pooled testing algorithms which are capable of large efficiency gains over ordinary pooled testing algorithms.


Since quantitative pooled testing is not our main focus in this chapter, we present here only a brief description of how a one-stage (nonadaptive) quantitative pooled testing algorithm works, in the context of testing for a disease. As in the previous section, we let $\mat A = (a_{ti})$ be the pooling matrix representing our pooling design, where $a_{ti} = 1$ if the sample from individual $i$ is included in the $t$th pooled test, and $a_{ti} = 0$ otherwise. We let $\tilde{\mat A}$ be the normalised pooling matrix, normalised so that all rows sum to one, i.e., $\tilde{a}_{ti} = 1/(\text{number of individuals in the }t\text{th pool})$ if the sample from individual $i$ is included in the $t$th pool, and $\tilde{a}_{ti} = 0$ otherwise. Unlike in the previous section, we now let $x_i$ denote the viral load in the sample from individual $i$; so that if individual $i$ is uninfected, then $x_i = 0$, while if individual $i$ is infected, then $x_i > 0$ (assuming that a sample was taken correctly from individual $i$). Our aim might be just to find the infected individuals (that is, to find the set of individuals $i$ such that $x_i > 0$), or we might want to estimate the actual viral loads, by estimating the whole vector $\vec x = (x_i)$. 

We assume that our pooled tests are capable of measuring the total viral load in each pool. 
For each $t$, we let $\tilde{y}_t$ denote the `true' total viral load in the $t$th pooled sample, by which we mean the value that we would report under perfect testing, without measurement errors. Then we have
\[ \tilde{y}_t = \sum_i \tilde{a}_{ti} x_i = (\tilde{\mat A} \vec x)_t .\]
Thus $\tilde{\vec y} = \tilde{\mat A} \vec x$ is given by a standard matrix--vector multiplication. The vector $\mathbf{y}$ of the actual measured results of the pooled tests will most likely differ from $\tilde{\vec y}$, due e.g.\ to measurement errors or procedural errors such as contamination. Ghosh et al \cite{tapestry1} propose an error model of multiplicative noise $y_t = \epsilon_t \tilde{y}_t$, where the $\epsilon_t$ are log-normally distributed, arguing that an additive error in the number of cycles corresponds to a multiplicative error in the estimate of the viral load. A more complex (but more realistic) error model would take into account the possibility of false positive results ($\tilde y_t = 0$ but $y_t > 0$, due e.g.\ to contamination) or false negative results ($\tilde y_t > 0$ but $y_t = 0$).

Our goal, given $\vec y$ and $\tilde{\mat A}$, is either to estimate $\vec x$, or perhaps just to find the coordinates of $\vec x$ with nonzero values. Mathematically, this is a special case of the well-studied \emph{compressed sensing} problem, on which there is extensive literature (which we will not attempt to summarize). The general aim of compressed sensing is to estimate (or guess) the value of an unknown vector $\mathbf x$ by taking measurements of a linear transform $\vec y = \tilde{\mat A} \vec{x}$ of $\vec x$, perhaps corrupted by noise, when $\mathbf x$ is known to be sparse (in the sense that many of its entries are $0$). This corresponds to our situation above in the setting of relatively low disease prevalence (the latter being generally the case, when pooled testing is under consideration for COVID-19). For a comprehensive introduction to the mathematics of compressed sensing, with emphasis on algorithms, we refer the reader to the book \cite{foucart}, and the references therein.

In the context of PCR testing for COVID-19, Ghosh et al \cite{tapestry1} propose the use of a scheme they call `Tapestry': a one-stage (nonadaptive) quantitative pooled testing algorithm that uses a range of $(r,s)$-regular designs with $r = 3$. In the early versions of the Tapestry algorithm, the total viral load in pooled samples was estimated using only the cycle threshold, but later versions of the algorithm use the intensity of fluorescence to obtain more accurate estimates of the total viral load \cite{manoj-personal} (see Section \ref{sec:testing-for-covid}, regarding cycle thresholds and fluorescence in PCR testing). Ghosh et al report that their protocol `has been validated with in vitro experiments that used synthetic RNA and DNA fragments', and that `validation with clinical samples is ongoing' \cite{tapestry1}.

\section{Practical challenges for pooled testing}
\label{sec:challenges}

The practical challenges and the downsides of implementing a pooled testing algorithm for COVID-19 testing -- either as part of a national or local testing programme 
or within an autonomous institution or company -- depend to some extent on the algorithm in question. A cost-benefit analysis is of course desirable, in each setting where pooled testing may be considered.

\myparagraph{Small benefit at high prevalence}
As mentioned earlier, when prevalence is greater than 38.2\%, no pooled testing algorithm  can outperform individual testing \cite{indiv} (under the assumption of perfect tests). Even when prevalence is significantly less than 38.2\%, it will often be judged that the down-sides of pooled testing outweigh the advantage of resource-savings. For example, pooled testing has been approved in India only for use in areas where the population prevalence is 2\% or less (see Section \ref{sec:brief-list}).

\myparagraph{Increased turnaround time}
As mentioned above, a single PCR test can typically be performed in four to six hours. If pooling can be automated, using e.g.\ a pipetting robot, then one-stage (non-adaptive) pooled testing algorithms will not have a significantly longer turnaround time than individual testing. For multistage pooled testing algorithms, such as Dorfman's algorithm, conservative two-stage pooled testing algorithms based on $(r,s)$-regular designs, or the multistage algorithm piloted in Rwanda (see Section \ref{sec:rwanda}), the increase in turnaround time (relative to individual testing) will depend partly on the laboratory set-up. The impact of increased turnaround times clearly depends upon the damage done by letting infections go undetected for longer; in the case of screening healthcare workers or social care workers, who work with individuals highly vulnerable to severe illness if exposed to SARS-CoV-2, this impact is likely to be much greater than in the case of screening university students or factory workers (for example).  

\myparagraph{Laboratory infrastructure}
Dorfman's algorithm does not require any sophisticated equipment to implement: the pooling of the samples can be done by hand. The pooling of the samples can even be done, as at the University of Cambridge (see Section \ref{sec:cam}), by the individuals to be tested, thus imposing no extra workload on laboratory staff. It only requires the laboratory to keep track of which individuals correspond to which pooled samples, and a capability to perform individual follow-up tests on the individuals whose pools test positive (or alternatively, for part-samples from each individual to be kept back during the first step, in case follow-up testing is needed on that individual in the second step). Some laboratories, e.g., those suffering from a shortage of well-trained personnel, or of equipment, would struggle to implement even Dorfman's algorithm (the simplest to implement): even keeping track of which samples belong to which individuals, once these have been divided in half, may prove challenging under conditions of extreme pressure and consequent disorganisation. 

The grid-algorithm is most efficiently implemented using a pipetting robot with an arm that can move in two dimensions; this piece of equipment, while commercially available, may be too expensive for organisations operating with a low budget, and for poorer countries. An alternative is to do the pooling manually, if sufficient manpower is available.

\myparagraph{The impact of dilution on test sensitivity} \label{sec:dilution}
When one infected sample is pooled with several others that are uninfected, the viral RNA is diluted, and this dilution leads to a decrease in the sensitivity of the PCR test on the pooled sample. The precise impact on sensitivity depends upon the laboratory protocol used, and also upon the distribution of viral loads in the samples being tested (which, in turn, depends upon the stage of the illness in the individuals being tested, as well as on individual biological factors). The following, however, gives a rough idea of the impact of dilution on sensitivity.
Using a common protocol, and a set of 838 SARS-CoV-2 positive specimens, Bateman et al \cite{bateman} found that dilution by a factor of five led (on average) to a 7\% reduction in sensitivity, dilution by a factor of ten led (on average) to a 9\% reduction in sensitivity, and dilution by a factor of 50 led (on average) to a 19\% reduction in sensitivity. By contrast, a systematic review of the accuracy of individual PCR tests found false negative rates ranging from 2\% to 29\%, using repeated PCR-testing as the gold standard (for true positivity). Using repeated PCR testing as the gold standard for positivity is likely to underestimate the true rate of false negatives.

More importantly, for tests taken in the field, is the fact that swabs are sometimes taken incorrectly. One community-based study of close contacts of confirmed COVID-19 cases in China, found an overall sensitivity of 71\% for upper-throat swabbing (by trained medical personnel) followed by an individual PCR test, using repeated PCR tests as the gold standard (for true positivity). When individuals self-swab, sensitivity is likely to be lower, unless the individuals themselves are appropriately trained (e.g., healthcare workers). Compared to these factors, overall, the impact of dilution on sensitivity, is relatively minor.

In the theoretical results we have described here, the number of samples per test $s$ can get extremely large when the prevalence $p$ is very small. In real applications, laboratories would be unwilling (due to dilution concerns) or unable (due to equipment capacity) to pool together very high numbers of samples. Thus for extremely low prevalences, the gains of pooled testing are unlikely to be as high as the theory suggests.

\myparagraph{How to deal with inconsistent test results}
When the sensitivity or the specificity of the tests being used is less than 100\%, a pooled testing algorithm can yield inconsistent results. For example, when Dorfman's algorithm is implemented, a particular pool can test positive due to the presence of one infected individual (a true positive), but then in the second stage of follow-up (individual) testing, all the individuals in that pool can test negative, due to the infected individual testing negative (a false negative). In this case, the testing authorities are faced with a dilemma: they could assume (wrongly, in this case), that the first (pooled test) was a false positive and that the follow-up tests were true negatives, or they could simply choose to declare the individuals in question free from infection (to avoid having to recall them for further testing), but they could also decide to repeat the second round of individual tests on that group (pool) of individuals, to hedge against the possibility that there was a false negative in the second round of individual tests. This could be regarded as a third, `confirmatory', testing-step in the algorithm. (If extra samples cannot be taken from the individuals in question, a third, confirmatory, testing-step may in fact be impossible.)

Which option the authorities choose, may depend on the impact of letting an infection go undetected; if the individual in question is a school pupil or a member of the general community (in a mass-testing programme), the impact will be less than if the individual is a healthcare worker working with patients highly vulnerable to severe illness or death in the case of COVID-19 infection, or if the individual is a resident-facing social care worker. The extra resource-requirements, and the delay, of a third (confirmatory) round of testing, should it be judged necessary, would have to be taken into account when deciding whether to adopt the pooled testing strategy.

\myparagraph{Prior prevalence estimates}
\label{sec:prevalence-estimates}
In situations where surveillance is poor, or infection levels are changing very rapidly, a laboratory may have very little idea of the prevalence of infection in an incoming batch of samples to be tested for SARS-CoV-2 infection. In such circumstances, a suboptimal choice of the parameters in a pooled testing algorithm algorithm (due to an underestimate of the prevalence) can lead to an inefficient second step -- less efficient, in fact, than individual testing. For example, if the prevalence is 25\%, then using Dorfman's algorithm with pools of size 10 requires on average approximately 1.04 tests per individual, which is slightly worse than individual testing (in addition to having a longer turnaround time). However, if an upper bound on the prevalence is known with a reasonably high degree of certainty (and this upper bound is not too high), then the parameters in a pooled testing algorithm can still be chosen so as to achieve a significant resource-saving over individual testing, even though the parameters cannot be fined-tuned to the exact prevalence. For example, the number of tests per individual for Dorfman's algorithm with fixed pool size $s$, is increasing in $p$, so the `worst case' is when $p$ is maximal. Hence, if the prevalence is known to be at most $1\%$, then Dorfman's algorithm with pools of size $10$, will require at most \(\tfrac{1}{10} + 1 - (1-p)^{10} \leq \tfrac{1}{10} + 1 - (1-0.01)^{10} \approx 0.196\) tests per individual (on average). Thus we get at least a five-fold improvement on individual testing, as long as the prevalence does not rise above $1\%$.
\nopagebreak

\myparagraph{Regulatory approval}
All of the issues listed above may be obstacles to regulatory approval. An additional obstacle to regulatory approval is the complexity of pooled testing algorithms, compared to individual testing. Often, policymakers will need to give their approval, bearing in mind public opinion, and a procedure which cannot be understood by a large percentage of the public (or by policymakers without the requisite quantitative training), may be less likely to gain such approval. On the other hand, the simpler pooled testing algorithms, such as Dorfman's algorithm and the grid algorithm, almost certainly can be explained in such a way that policymakers (and the majority of the public) can understand them -- and this may well be part of the reason why these two algorithms have seen the widest use, of all pooled testing algorithms, during the COVID-19 pandemic so far.

\section{Uses of pooled testing in the COVID-19 pandemic}
\label{sec:uses}

Hitherto in the \covid\ pandemic, Dorfman's algorithm has been the most widely-used pooled testing strategy. This is almost certainly because it is (i) easy to implement without necessarily needing large changes in laboratory equipment or infrastructure, (ii) relatively robust to changes in prevalence, (iii) simple and transparent enough for non-scientific decision makers or the public to understand and for regulators to approve, and (iv) of easily predictable and controllable sensitivity. At the same time, it can yield large efficiency gains over individual testing, as outlined above. In this section, we give some examples of places where pooled testing has been applied during the COVID-19 pandemic, focussing on examples where detailed information is available.

\subsection{Dorfman's algorithm at the University of Cambridge}
\label{sec:cam}

Starting on 6 October 2020, University of Cambridge students in college accommodation (approximately 16,000 students) were asked to participate in the University's asymptomatic COVID-19 screening programme, which was based on Dorfman's algorithm \cite{cam}.

In detail, students in college accommodation were grouped into `social bubbles' of average size 8 (and maximum size 10), according to shared living facilities (e.g., a bathroom, kitchen or living room). Students were asked not to socialise outside their social bubbles. In the asymptomatic screening programme, students in each social bubble were asked to take swabs once per fortnight (increasing in December to once per week), and to combine them into a single container. The container was then sent to a local laboratory for pooled PCR-testing. When the pooled test was complete, all the students in the corresponding pool were informed of the result (usually by text message, the evening of the day the swabs were sent off), and if the pooled test was positive, all students in the social bubble were instructed to self-isolate immediately, and to take an individual test as soon as possible (and if possible, before midday the next day at the latest), at a national testing site. If at least one student in the bubble tested positive in their individual follow-up test, all students in bubble were instructed to continue their self-isolation. If all students in the bubble tested negative in their individual follow-up tests, the pooled test was assumed to be a false positive, and all students in bubble were released from self-isolation.

Participation was voluntary, but consent rates were high, starting at 75\% 
in the first week of the Autumn term, and increasing to 82\% 
in the last week of the Autumn term. Symptomatic students were excluded, as they were instructed to seek an individual test at a national testing site as soon as possible after experiencing symptoms. Students who had recently tested PCR-positive were also excluded. The total turnaround time (of pooled testing plus follow-up individual test) was typically around 48 hours when follow-up testing was done, but students were typically informed of a positive pooled test result 
approximately eight hours after sending off the swabs. A simple rota determined which students were asked to provide swabs on which weeks; by the final two weeks of the Autumn term, all the students in each bubble were being asked to provide swabs simultaneously, once per week.

The numbers of positive cases detected each week by this programme are shown in Table 5.

\newcommand{\specialcell}[2][c]{%
  \begin{tabular}[#1]{@{}c@{}}#2\end{tabular}}

\begin{table}[!t]
\caption{Results of screening participating (and asymptomatic) University of Cambridge students living in College accommodation, using Dorfman's algorithm with pools of size up to 10, from 5 October to 6 December 2020.}\label{table:ucam} \label{tab:results-cam}
\begin{center}
\begin{tabular}{lccc}
\hline
Week & \# screened & \specialcell{\# confirmed \\[-6px] positive cases}  & \specialcell{\% confirmed \\[-6px] positive cases} \\
\hline
5--11 Oct & 3,463 & 12 & 0.3\% \\
12--18 Oct & 3,675 & 34 & 0.9\% \\
19--25 Oct & 4,660 & 39 & 0.8\%\\
26 Oct--1 Nov & 5,494 & 38 & 0.7\% \\
2--8 Nov & 4,583 & 23 & 0.5\% \\
9--15 Nov & 5,339 & 80 & 1.5\% \\
16--22 Nov & 4,138 & 27 & 0.7\% \\
23--28 Nov & 9,329 & 3 & 0.03\% \\
29 Nov--6 Dec & 9,376 & 0 & 0\% \\
\hline
\end{tabular}
\end{center}
\end{table}

We now discuss, briefly, the resource requirements of this scheme.
In the last two weeks of the Autumn term, all the students in each bubble were asked to provide swabs. For example, in the eighth week of term (23--29 November), 1,927 pooled tests were carried out, and 9,481 students contributed swabs; seven pools tested positive \cite{week8}, requiring between 7 and 70 follow-up individual tests. Hence, a total of between 1,934 and 1,997 PCR tests were required in the eighth week of the Autumn term, i.e., either 20\% or 21\% of the number of PCR tests that would have been required to individually test the 9,481 students who contributed swabs.

As mentioned above, students were requested not to socialise outside their bubbles. Assuming a high level of compliance, this would mean that positive cases were more likely to cluster within bubbles, leading to a resource-saving at the second stage (of follow-up individual testing): this is the `best case' in Dorfman's algorithm, in terms of resource-use at the second step, when the infected individuals are distributed among as few pools as possible. In the sixth week of term, for example, 80 students individually tested positive across 59 positive-testing pools \cite{week6}, giving an average of approximately $1.3$ infected students per positive testing pool. Had the infected individuals been more evenly distributed, there would have been only one positive-testing student per positive-testing pool.

\subsection{The grid and P-BEST algorithms in Israel} \label{sec:israel}

In August 2020, Israel's Ministry of Health approved two single-step pooled testing protocols for use in clinical laboratories in the country: one based upon the `grid algorithm', and one based upon the `P-BEST' algorithm of Shental et al \cite{shental}. The grid algorithm has been described earlier.

P-BEST uses an $(r = 6, s = 48)$-regular design with a code-based construction. This deals with $N = 384$ individuals in $T = Nr/s = 48$ pooled tests. A `best guess' for the identifies of the infected individuals (from the results of the pooled tests), is obtained using the method described in Section \ref{onestage}, above.

It should be noted that the P-BEST algorithm does require software and computing resources to implement, in addition to a pipetting robot with an arm that can move in two dimensions. However, this equipment is affordable by most countries.

In trials where at most 5 out of 384 individuals are PCR-positive under individual PCR testing (corresponding to a prevalence of 1.3\% or lower), the P-BEST algorithm usually correctly identified all infected and uninfected individuals. Problems can arise when a batch of samples is received with a much higher prevalence than 1.3\%; in this case, even if there are no false positive and or false negative test results, it is often not possible to use P-BEST to determine which individuals are infected and which are not.

In Israel, in clinical laboratories where P-BEST (or the grid algorithm) has been employed, data analysis and machine-learning has been employed also, to predict which batches of samples are likely to have much higher prevalence rates than the national average (based on origin); such batches were typically dealt with using individual testing.

\subsection{A multi-stage $(r,s)$-regular algorithm in Rwanda}
\label{sec:rwanda}

Starting in August 2020, a multi-stage algorithm was piloted in  Rwanda, where infection prevalence was low but the supply of PCR tests was limited \cite{turok}.

The stages are as follows. It is required to pick two integer parameters, $a$ and $r_2$.
\begin{enumerate}
\item A Dorfman-like stage with $r_1 = 1$ pooled test for each individual and $s_1 = a^{r_2}$ samples in each test. A negative result shows that all $s_1$ individuals are uninfected; individuals in a positive pool go through to stage two.
\item An $(r_2, s_2 = a^{r_2-1})$-regular design, with a hypercube construction (see Section \ref{sec:rs1}).
\item If the hypercube contains zero or one infected individuals, they can be identified. Otherwise, further stages of testing are used to disambiguate the results; we don't go into details here.
\end{enumerate}

The parameters are chosen to be as efficient as possible while still ensuring that the chance of more than two stages being required is very small.
One common choice is $a = 3, r_2 =3$, so that the first stage is a $(1, 81)$-regular Dorfman-like design, and the second stage is a $(3, 9)$-regular hypercube design of 9 tests for 81 individuals.

\subsection{Other uses of pooled testing}
\label{sec:brief-list}
Here is a brief (and far from exhaustive) list of some other examples of the use of pooled testing in the \covid\ pandemic.

\begin{itemize}
\item PCR testing using Dorfman's algorithm in Wuhan, China \cite{wsj}. Between 12 May and 1 June 2020, 9.9 million Wuhan residents were tested; the vast majority were asymptomatic. Dorfman's algorithm was reportedly used for approximately 25\% of this testing, with pools of sizes between 5 and 10. Only 300 positive cases were identified. For the approximately 25\% of residents tested using the pooled testing (Dorfman) method, pooled testing would have used between one fifth and one tenth (approximately) of the number of PCR-tests that would have been required by individual testing, at this prevalence level. (Whether resource-use was closer to one fifth or to one tenth, would depend upon the most common pool-size used; we are not aware that data on this has been publicly released.)
 \item Screening of students on University campuses, using Dorfman's algorithm or the grid algorithm, during the Autumn/Fall term of 2020: Universit\'e de Li\`ege, Belgium (saliva samples using Dorfman's algorithm with pools of size 8) \cite{liege} ; Duke University, USA (saliva samples using Dorfman's algorithm with pools of size between 5 and 10; participation mandatory for students on campus) \cite{duke} ; Michigan State University, USA (saliva samples using a grid algorithm) \cite{msu} ; Syracuse University, USA (saliva samples using Dorfman's algorithm with pools of size between 20 and 25) \cite{syracuse} ; Shenandoah University, (saliva samples using Dorfman's algorithm with pools of size 4 or 5) USA \cite{shenandoah} .
\item PCR testing using Dorfman's algorithm with pools of size up to 20, by Fundaci\'on Biom\'edica Galicia Sur, Galicia, Spain, to screen asymptomatic healthcare workers, social care workers, industrial workers and port workers in the province of Galicia from September 2020 onwards, raising screening-capacity to 100,000 screenings per month (with health and social care workers being screened twice per week). Pipetting robots have been used. \cite{galicia}
    \item PCR testing using Dorfman's algorithm with pools of size up to 30, by Saarland University Hospital, Germany, for the regular screening of asymptomatic hosptial patients and hospital staff, and care home residents in Saarland, from March 2020 onwards. Approximately 22,000 people screened. \cite{saarland}
    \item PCR testing using Dorfman's algorithm with pools of size 10, by Noguchi Memorial Institute for Medical Research, Ghana, to test contacts of confirmed cases. Initially 10,000 people tested per day, from April 2020 onwards. \cite{who-ghana}
    \item PCR testing using Dorfman's algorithm with pools of size 5, by the states of Uttar Pradesh \cite{up} and West Bengal \cite{wb}, India, in areas with estimated prevalence of 2\% or lower.
\end{itemize}

\section{Applications of pooled testing for COVID-19: some conclusions}
\label{sec:personal-recommendations}

In this section, we conclude, by drawing some of the above analysis together and discussing our own personal perspective on the practical settings where pooled testing for COVID-19 is likely to be useful, or at least may merit serious consideration.

\subsection{Pooled testing for asymptomatic subpopulations}

As stated in the introduction, we believe that pooled testing is most likely to be useful for the screening of asymptomatic people, for surveillance, and possibly for the testing of contacts of confirmed cases, provided the prevalence of infection among the group to be tested is sufficiently low. On the other hand, we believe that in most countries, pooled testing is unlikely to be useful for the testing of symptomatic people. (Here, we use the term `asymptomatic' to denote someone who is not experiencing the recognised symptoms 
at the time of their test. This includes `true asymptomatics', who never experience symptoms, and `pre-symptomatics', who go on to develop symptoms after their test.)

There are two main reasons why we believe that, in most countries, pooled testing will be of limited use for the testing of symptomatic people. First, the prevalence of COVID-19 infection among those presenting symptoms is usually sufficiently high that the resource savings of pooled testing are modest compared to individual testing, and may be outweighed by the down-sides of pooled testing, such as increased turnaround time. Among those presenting COVID-like symptoms, prevalences of between $4\%$ and $33\%$ are realistic, depending upon the setting, the location, the symptoms used in the definition of `symptomatic', and the prevalence of other respiratory viruses (which in turn depends on the time of year) \cite{kcl, pueyo}. Second, many countries already have well-established testing programmes using individual testing for those presenting symptoms. In many countries, including the UK, an individual test on symptomatic people is mandated by the regulatory authorities to confirm infection, even if they can be proved positive solely via pooled tests.

On the other hand, in many countries, the prevalence of COVID-19 infection among the general population has for quite long periods been at levels low enough that pooled testing of asymptomatic people can yield large efficiency gains over individual testing. For example, the estimated prevalence of current \sars\ infection in the general community 
in England, as estimated by the ONS Infection Survey \cite{ons-infection-survey}, has ranged from 0.026\% in early July, to 2.1\% in early January (and 3.6\% in London in early January). Hence, for the prevalence among asymptomatics in England, values of $p$ between $0.02\%$ and $1.6\%$ are good estimates. Countries that have adopted more stringent non-pharmaceutical interventions (such as stricter lockdowns or strongly enforced quarantines for international arrivals)
have experienced lower prevalence rates; for example, New Zealand probably eliminated \covid\ infections in the general community between early May and mid-August 2020, except for international arrivals, who were quarantined \cite{nz-nejm}.

As mentioned above, in most countries with substantial COVID-19 testing resources, symptomatic individuals are instructed to seek an individual test, so if they are representative of the general population of asymptomatic people, a subpopulation to be screened (possibly using pooled testing) is likely to have a slightly lower prevalence of infection than that among the whole population (which obviously includes symptomatic people). For example, the UK's Office for National Statistics estimated that in early October 2020, 24\% of infected individuals were symptomatic 
on the day of their positive test \cite{ons-asymp}, so one could expect, absent significant changes to the proportion of asymptomatic infections,
that the prevalence among an asymptomatic group of people (if representative of the population) would about 24\% less than the prevalence among the whole population. We note that, if/when the relative prevalence of a new variant becomes more common, the proportion of asymptomatic infections can be expected to change. For example, according to the ONS Infection Survey \cite{ons-jan}, the so-called Kent (or `British') variant (Variant of Concern 202012/01) produces a greater proportion of symptomatic infections than wild-type SARS-CoV-2.

It is often desirable to screen subpopulations where the prevalence of infection is likely to be significantly higher than in the general population: for example, patient-facing healthcare workers, resident-facing social care workers, and factory workers in high-risk environments such as meat-processing plants.
But itt may also be desirable to screen subpopulations where the prevalence of infection is likely to be similar to the general population: see some of the examples below.

We list here some of the settings where we believe pooled testing for COVID-19 is most likely to be useful. Of course, a careful cost-benefit analysis should be carried out for each potential application, with the decision to adopt or not depending on certain factors, including the prevalence level, laboratory resources and capabilities, the impact of increasing the turnaround time, and regulatory constraints.

\begin{itemize}
    \item {\bf University students.} As may be apparent from the relatively large number of examples of this in Section \ref{sec:uses}, screening of asymptomatic university students is one of the less controversial applications of pooled testing. Severe illness is very rare among those of student age, so the impact on students of an increase in the turnaround time associated with pooled testing  is slight. If a significant amount of in-person teaching is taking place, there are higher risks to older members of staff in the event that they are infected, so the impact on them of increased turnaround time should be taken into account.
    \item {\bf Key workers} -- for example factory workers, warehouse workers and port workers (but excluding patient-facing healthcare workers and resident-facing social care workers) provided the prevalence is not too high among the workers in question. We recall from Section \ref{sec:brief-list} that pooled testing has been used in Galicia for the screening of factory and port workers.
    \item {\bf Members of sports teams}. Screening of asymptomatic members of sports teams is suggested in \cite{turok}.
    \item {\bf Airline passengers}. This is also suggested in \cite{turok}.
    \item {\bf Non-household contacts of confirmed cases}, provided the estimated prevalence among these is sufficiently low. We recall from Section \ref{sec:brief-list} that this has been done in Ghana.
\end{itemize}

We also mention two settings where we are more doubtful of the appropriateness of pooled testing of asymptomatic people: school pupils, and health and social care workers.

\paragraph{School pupils}

We are not aware of any countries that have carried out regular screening of asymptomatic school pupils, using pooled testing on a large scale (though there have been some local research projects using pooled testing for screening in schools, for research purposes, e.g.\ \cite{commins}).

The logistical challenges of regular screening of school pupils are much greater than in the case of university students, who can (if necessary) organise much of the process themselves (see Section \ref{sec:cam}). School pupils cannot do this, so if such screening is to be done, an additional organisational burden would be placed on schools, which are already overstretched and (in the UK) are required to help children catch up on months of missed in-person teaching \cite{dfe-catch-up}. Extra funding, more space, and additional staffing would be required for screening in schools. A further problem is that there is little incentive for low-income families to give permission for their children to participate in regular screening, since if a child tests positive and this is reported, the parent or guardian will have to take time off work. (Financial compensation for this might help to incentivise low-income parents to give such permission.)  Primary school aged children are unlikely to tolerate regular swabbing (see e.g.\ the advice from Public Health England reported in \cite{rapid-asym}), so the question is more relevant for secondary school pupils -- at least unless/until reliable saliva tests become widely available.

To maximise the infection-control benefit of screening school pupils, school contacts of positive-testing pupils would be asked to self-isolate. In the UKm for example, since the beginning of the Autumn Term in September 2020, school pupils have been grouped together into `bubbles', with limited mixing between bubbles, but much more mixing within bubbles. In UK secondary schools, these bubbles have often been the size of year-groups or half-year-groups, due to timetabling and space constraints \cite{dfe-bubbles}. This means that in the event of a pupil testing positive for COVID-19, secondary schools (and the local public health bodies advising them) are often faced with a difficult choice: either the pupil's entire bubble is requested to self-isolate, leading to more missed in-person teaching, or there is a decision to accept the risk that onward transmission will take place in the bubble (unless the pupil's closest contacts within the bubble can be reliably traced, which is not always possible). The self-isolation of entire bubbles during the Autumn Term, combined with earlier lockdowns and school closures, has already led to approximately 110 school days lost from in-person teaching, for the average pupil, from March 2020 to March 2021 \cite{guardian-exeter} (out of a total of 190 school days in a usual school year). Regular screening would increase the frequency of entire bubbles having to self-isolate, both due to false positives and due to true positives.

We believe that the screening of school pupils should only be adopted if and when the benefit (in terms of reduced community transmission, and disease burden) is clear. School-aged children are extremely unlikely to suffer severe disease in the event of contracting COVID-19. (Of the 12 million under-18s in England during 2020, only about one in 2000 were admitted to hospital due to COVID-19, and of these, less than one in 50,000 were admitted to intensive care. For PIMS-TS, a very rare but severe inflammatory condition associated with previous SARS-CoV-2 infection in children, about one in 17,000 under-18s in England were admitted to hospital during 2020, and less than one in 40,000 admitted to intensive care \cite{viner}.) An analysis by the UK's Office for National Statistics found no statistically significant difference between infection-rates in primary and secondary-school teachers versus the average infection-rate  between 2 September and 16 October 2020 \cite{ons-teachers}. Moreover, a systematic review \cite{review} of low-bias observational studies concluded that, as of March 2021, schools had not been a major driver of community transmission. (This may change as new variants arise and vaccination coverage increases among adults, so the evidence should be continually monitored.) On the other hand, the long-term negative economic, social and health impacts of missed schooling are well-documented \cite{delve}, with a missed year of schooling causing on average a reduction of 8\% in lifetime earnings \cite{edu-returns}. The impact of missed in-person teaching is likely to be particularly severe for children from disadvantaged backgrounds, who are likely to be much less able to access virtual learning resources -- due to lacking a quiet study space, IT equipment, or a conducive home environment for learning \cite{delve,montacute}.

In the event of schools becoming a major driver of community transmission in the future (and assuming such community transmission remains a problem, due e.g.\ to the continued absence of highly-effective treatments, and the imperfect level of protection afforded by the available vaccines, to the vulnerable), we believe that the vaccination of schoolchildren would be preferable to their regular screening --- yielding greater benefits in terms of reduced community transmission, and imposing far less disruption on the lives of children, though we await more data on the side-effects of the currently available vaccines in children, to be sure of this. (We note that in the US, most 12-15 year-olds are already eligible to receive the Pfizer vaccine, as of May 2021 \cite{nyt-may}.)

\paragraph{Health and social care workers}

One potential application of pooled testing that many authors are circumspect about is the screening of asymptomatic health and social care workers. Given the major risks to vulnerable patients and social care residents associated with any additional delay in finding positive cases among these workers, screening using individual testing is generally agreed to be preferable, if there are sufficient resources. Even in the presence of very severe resource constraints, a careful cost-benefit analysis should be performed to compare the impact of screening based on individual testing with that based on pooled testing, taking into account the increased turnaround time. Even Mutesa et al \cite{turok}, who are in general strong advocates of the use of pooled testing for COVID-19, state explicitly that they do not advocate its use for the screening of healthcare workers. We do note, however, that pooled testing was used by Saarland University Hospital for this purpose (see Section \ref{sec:brief-list}).

\subsection{Pooled testing and vaccination programmes}

At the time of writing (March 2021), vaccination programmes are proceeding rapidly in many developed countries, and are having a large effect in reducing the number of COVID-19 cases. It might be thought that, in such countries, there will soon be no need to consider pooled testing. We believe that such an assumption may be premature at this stage, mainly because we do not yet have reliable data on the extent to which the vaccines currently being distributed reduce the number of cases (symptomatic and asymptomatic) of new variants of COVID-19, particularly the South African and Brazilian variants \cite{bmj-variants-vaccines}. We proceed to give an outline of the current evidence-base on vaccine effectiveness (as we understand it), at the time of writing (March 2021).

Preliminary data relating to impact of the Pfizer/BioNTech vaccine in Israel (the country with the second-most advanced vaccination programme, which has also released the most data), is very promising, with Haas et al \cite{haas} reporting an estimated vaccine effectiveness of 95.3\% (95\% CI: 94.9\% - 95.7\%) against SARS-CoV-2 infection (either symptomatic or asymptomatic infection), from seven days after the administration of the second dose, based upon national surveillance of the more than 4.7 million vaccinated individuals.

However, there is reason to believe that current vaccines will be somewhat less effective against new variants. For example, early results from trials in South Africa conducted by Madhi et al \cite{madhi} suggest that the Oxford/AstraZeneca vaccine provides much less protection against mild to moderate disease caused by the South African variant than against that caused by `wild-type' SARS-CoV-2. However, it is also believed that it will not take very long to adapt the Oxford/Astra-Zeneca vaccine to work better against the South African variant, and work is underway to that effect, with 
the leader of the Oxford team, stating on 7 February 2021 that `it looks very much like [the adapted vaccine] will be available for the Autumn' \cite{gilbert-guardian}. We note that the South African variant is probably not yet in very wide circulation in Israel, with a few hundred confirmed active cases of the South African variant on 19 February 2021, compared to 48,018 confirmed active cases overall \cite{toi} -- although the former is probably an underestimate, due to constraints on genome sequencing capacity (which is needed to confirm the presence of the South African variant). Hence, the highly positive results of the Pfizer/BioNTech vaccine there do not imply equally high effectiveness against the South African variant. A very recent announcement by the Israeli Ministry of Health, according to the {\em Times of Israel}, stated that an early study suggests the Pfizer/BioNTech is six times less effective against the South African variant than against `wild-type' SARS-CoV-2 \cite{toi}, though the numbers of people involved in the study was not mentioned; we are not aware that further details have been released at this stage.

As far as we are aware, reliable data is not yet available on the effectiveness of the vaccines currently being distributed, against the Brazilian variant \cite{bmj-variants-vaccines}.

Bearing in mind this uncertainty, and the risk of further new variants arising that are resistant to available vaccines, we believe it would be wise for decision-makers to bear in mind the possibility that it may be desirable to rapidly increase testing-capacity using pooled testing in the medium term. In poorer countries, vaccination programmes are likely to be long delayed. In such countries, pooled testing may still be a valuable tool to consider for the foreseeable future.

\subsection{Pooled testing for surveillance}

At low prevalence levels, pooled testing has the potential for very large resource-saving in national COVID-19 surveillance programmes. In some countries with a very large testing capacity, this may not be necessary -- for example, the UK's ONS (Coronavirus) Infection Survey currently tests a random sample of approximately 400,000 members of the community population in England, once per fortnight, using individual testing \cite{ons-meth}; compared to the UK's Pillar II testing capacity of more than 200,000 tests per day, this is not too great a resource requirement. However, if it is desired to reduce the resource requirements of a nationwide surveillance programme, pooled testing provides a way of doing so. Using pools of size up to 100, the hypercube-based algorithm piloted in Rwanda by Mutesa et al \cite{turok} can estimate the prevalence fairly accurately, while achieving an approximately 100-fold reduction in number the tests used when the prevalence level is at $0.05\%$. The main concern here is the reduction in sensitivity caused by dilution, but Mutesa et al \cite{turok} report proof-of-concept experiments which suggest that, using an appropriate protocol, sensitivities of 98\% or 92\% (depending on the gene targeted), can be achieved at a 100-fold level of dilution. This suggests that their scheme can be used reliably to monitor prevalence. The extra turnaround time compared to individual testing is likely to be much less of an issue with surveillance than with case identification, particularly at low prevalence levels.

It is also plausible that pooled testing could be used to monitor the prevalence of new variants. This may become particularly important if new variants begin to seriously hinder the success of vaccination programmes. A new PCR-testing method (involving only a minor update to existing PCR tests) that can detect which variant of SARS-CoV-2 a patient is carrying is currently undergoing clinical trials by the biotechnology firm Novozymes \cite{ns-novo}.

\section*{Acknowledgements}
The authors were supported in part by UKRI Research Grant EP/W000032/1. The authors thank Julia Eisenberg, Carmen Boado-Penas and \c{S}ule \c{S}ahin for helpful comments.

\end{document}